\documentclass[twocolumn]{autart}    
\usepackage[cmex10]{amsmath}
\numberwithin{equation}{section}
\usepackage{amssymb}
\usepackage{mathdots}
\usepackage{psfrag}
\usepackage{verbatim}
\usepackage{nameref}

\usepackage{lineno,hyperref}
\modulolinenumbers[5]
\usepackage{natbib}
\newcommand{\overbar}[1]{\mkern 1.5mu\overline{\mkern-1.5mu#1\mkern-1.5mu}\mkern 1.5mu}

\usepackage{mathrsfs}
\usepackage{mathtools}
\usepackage{mathdots}
\usepackage{graphicx}          
\usepackage{wrapfig}

\usepackage{enumitem}
\newcounter{muni}
\newenvironment{remunerate}
               {\begin{list}{{\upshape 
               \arabic{muni}.}}{\usecounter{muni}
                \setlength{\leftmargin}{0pt}
                \setlength{\itemindent}{25pt}}}{\end{list}}

\makeatletter
\newcommand{\labitem}[2]{%
\def\@itemlabel{#1}
\item
\def\@currentlabel{#1}\label{#2}}
\makeatother

\begin{document}

\begin{frontmatter}

\title{A theory of passive linear systems with no assumptions}

\thanks[footnoteinfo]
{A simpler version of Theorem \ref{thm:pbt} in this paper, for single-input single-output systems, was presented at the European Control Conference, Aalborg, 2016 \citep[see][]{Hugcsisop}.}
\thanks[footnoteinfo2]
{\copyright \hspace{0.03cm} 2017. This manuscript version is made available under the CC-BY-NC-ND 4.0 license http://creativecommons.org/licenses/by-nc-nd/4.0/. This is the accepted version of the manuscript: Hughes, T.H.: A theory of passive linear systems with no assumptions, Automatica, 86, 87-97 (2017).}

\author[thh22]{Timothy H.\ Hughes \thanksref{footnoteinfo}\thanksref{footnoteinfo2}}\ead{t.h.hughes@exeter.ac.uk}   
\address[thh22]{Department of Mathematics, University of Exeter, Penryn Campus, Penryn, Cornwall, TR10 9EZ, UK}  
          
\begin{keyword}                           
Passive system; Positive-real lemma; Linear system; Controllability; Observability; Behavior. 
\end{keyword}                             

\begin{abstract}                          
We present two linked theorems on passivity: the passive behavior theorem, parts 1 and 2. Part 1 provides necessary and sufficient conditions for a general linear system, described by a set of high order differential equations, to be passive. Part 2 extends the positive-real lemma to include uncontrollable and unobservable state-space systems. 
\end{abstract}

\end{frontmatter}

\newtheorem{example}[thm]{Example}

\section{Introduction}
\label{sec:intro}
A system is called \emph{passive} if there is an upper bound on the net energy that can be extracted from the system from the present time onwards. This is a fundamental property of many physical systems. In systems and control theory, the concept of passivity has its origins in the study of electric networks comprising resistors, inductors, capacitors, transformers, and gyrators (RLCTG networks). In contemporary systems theory, passive systems are more familiar through their role in the positive-real lemma. This lemma proves the equivalence of: (i) an integral condition related to the energy exchanged with the system; (ii) a condition on the transfer function for the system (the positive-real condition); and (iii) a linear matrix inequality involving the matrices in a state-space realization for the system. As well as being relevant to passive systems, the lemma also gives necessary and sufficient conditions for the existence of non-negative definite solutions to an important linear matrix inequality and algebraic Riccati equation, and has links with spectral factorisation. However, these results are all subject to one caveat: the system is assumed to be controllable. 

As emphasised by \citet{camwb, JWDDS, HugNa}, there is no explicit connection between the concepts of passivity and controllability. Moreover, the a-priori assumption of controllability in the positive-real lemma leaves open several questions of physical significance. In particular, it is not known what uncontrollable behaviors can be realized as the driving-point behavior of an electric (RLCTG) network. Similarly, necessary and sufficient conditions for the existence of a non-negative definite solution to the linear matrix inequality (and algebraic Riccati equation) considered in the positive-real lemma are unknown when the state-space realization under consideration is uncontrollable. There have been many papers in the literature that have aimed to relax the assumption of controllability in the positive-real lemma, e.g., \citet{POPRL, CLJKYPL, KPRLNMR} (and many papers have studied uncontrollable cyclo-dissipative systems, e.g., \citet{FPPRLE, camwb, FPRLVNA, PBDUS}), but all of these papers contain other a-priori assumptions. The objective of this paper is to provide a complete theory of passive linear systems with no superfluous assumptions. Our main contributions are: 1.\ a new trajectory-based definition of passivity (Definition \ref{def:pb}); and 2.\ two linked theorems that we call the \emph{passive behavior theorem}, parts 1 and 2. Part 1 (Theorem \ref{thm:pbtp2}) provides necessary and sufficient conditions for the passivity of a general linear system (described by a differential equation of the form $P(\tfrac{d}{dt})\mathbf{i} = Q(\tfrac{d}{dt})\mathbf{v}$ for some square polynomial matrices $P$ and $Q$). This generalizes classical results that are restricted to controllable behaviors (where $P$ and $Q$ are left coprime). Part 2 (Theorem \ref{thm:pbt}) extends the positive-real lemma by removing the a-priori controllability and observability assumptions. As a corollary of these results, we find that any passive (not necessarily controllable) behavior can be realized as the driving-point behavior of an electric (RLCTG) network

The structure of the paper is as follows. In Section \ref{sec:pbaprl}, we discuss the positive-real lemma and its limitations. Section \ref{sec:pass} discusses our new definition of \emph{passivity}. Then, in Section \ref{sec:pbts}, we introduce the new concept of a \emph{positive-real pair}, and we state our two passive behavior theorems. It is shown that our new concept of a positive-real pair provides the appropriate extension of the positive-real concept to uncontrollable systems. Specifically, for any pair of square polynomial matrices $P$ and $Q$, we show that the system corresponding to the solutions to the differential equation $P(\tfrac{d}{dt})\mathbf{i} = Q(\tfrac{d}{dt})\mathbf{v}$ is passive if and only if $(P, Q)$ is a positive-real pair. The proofs of the passive behavior theorems are in Section \ref{sec:pbtp}, and some preliminary results appear in Section \ref{sec:pbprp}. Finally, the paper is strongly influenced by the behavioral approach to dynamical systems \citep[see][]{JWIMTSC}. Therefore, to make the paper accessible to the reader unfamiliar with behavioral theory, we provide four short appendices containing relevant background on linear systems, behaviors, and polynomial matrices. These contain numbered notes (A1, A2, and so forth) that will be referred to in the text. The reader who wishes to follow the proofs in Sections \ref{sec:pbprp} and \ref{sec:pbtp} is advised to first read these appendices.

The notation is as follows. $\mathbb{R}$ ($\mathbb{C}$)  denotes the real (complex) numbers; $\mathbb{C}_{+}$ ($\overbar{\mathbb{C}}_{+}$) denotes the open (closed) right-half plane; $\mathbb{C}_{-}$ ($\overbar{\mathbb{C}}_{-}$) denotes the open (closed) left-half plane. $\mathbb{R}[\xi]$ ($\mathbb{R}(\xi)$) denotes the polynomials (rational functions) in the indeterminate $\xi$ with real coefficients. $\mathbb{R}^{m \times n}$ (resp., $\mathbb{C}^{m \times n}, \mathbb{R}^{m \times n}[\xi], \mathbb{R}^{m \times n}(\xi)$) denotes the matrices with $m$ rows and $n$ columns with entries from $\mathbb{R}$ (resp., $\mathbb{C}, \mathbb{R}[\xi], \mathbb{R}(\xi)$), and the number $n$ is omitted whenever $n=1$. If $H \in \mathbb{C}^{m \times n}$, then $\Re{(H)}$ ($\Im{(H)}$) denotes its real (imaginary) part, and $\bar{H}$ its complex conjugate. If $H \in \mathbb{R}^{m \times n}, \mathbb{C}^{m \times n}, \mathbb{R}^{m \times n}[\xi]$ or $\mathbb{R}^{m \times n}(\xi)$, then $H^{T}$ denotes its transpose; and if $H$ is nonsingular (i.e., $\det(H) \not\equiv 0$), then $H^{-1}$ denotes its inverse. We let $\text{col}(H_{1} \hspace{0.15cm} \cdots \hspace{0.15cm} H_{n})$ ($\text{diag}(H_{1} \hspace{0.15cm} \cdots \hspace{0.15cm} H_{n})$) denote the block column (block diagonal) matrix with entries $H_{1}, \ldots , H_{n}$. If $M \in \mathbb{C}^{m \times m}$, then $M > 0$ ($M \geq 0$) indicates that $M$ is Hermitian positive (non-negative) definite, and $\text{spec}(M) \coloneqq \lbrace \lambda \in \mathbb{C} \mid \text{det}(\lambda I {-} M) = 0\rbrace$. If $G \in \mathbb{R}^{m \times n}(\xi)$, then $\text{normalrank}(G) \coloneqq \max_{\lambda \in \mathbb{C}}(\text{rank}(G(\lambda)))$, $G^{\star}(\xi) \coloneqq G(-\xi)^{T}$, $G$ is called \emph{para-Hermitian} if $G = G^{\star}$, and \emph{proper} if $\lim_{\xi \rightarrow \infty}(G(\xi))$ exists. $\mathcal{L}_{1}^{\text{loc}}\left(\mathbb{R}, \mathbb{R}^{k}\right)$ and $\mathcal{C}_{\infty}\left(\mathbb{R}, \mathbb{R}^{k}\right)$ denote the ($k$-vector-valued) locally integrable and infinitely-often differentiable functions \citep[Definitions 2.3.3, 2.3.4]{JWIMTSC}. We equate any two locally integrable functions that differ only on a set of measure zero. If $\mathbf{w} \in \mathcal{L}_{1}^{\text{loc}}\left(\mathbb{R}, \mathbb{R}^{k}\right)$, then $\mathbf{w}^{T}$ denotes the function satisfying $\mathbf{w}^{T}(t) = \mathbf{w}(t)^{T}$ for all $t \in \mathbb{R}$. We also consider the function space
\begin{align*}
&\hspace*{-0.3cm} \mathcal{E}_{\mathbb{C}_{-}}\!\left(\mathbb{R}, \mathbb{R}^{k}\right)\! {\coloneqq} \lbrace \mathbf{w} \mid \mathbf{w}(t) {=} \Re{\left(\!\sum_{i=1}^{N}\sum_{j=0}^{n_{i}-1}\tilde{\mathbf{w}}_{ij}t^{j}e^{\lambda_{i}t}\!\right)} \text{ for all } t {\in} \mathbb{R} \nonumber \\  &\hspace{2.0cm} \text{ with } \tilde{\mathbf{w}}_{ij} \in \mathbb{C}^{k}, \lambda_{i} \in \mathbb{C}_{-}, \text{ and }  N, n_{i} \text{ integers} \rbrace, 
\end{align*}
and note that $\mathcal{E}_{\mathbb{C}_{-}}\!\left(\mathbb{R}, \mathbb{R}^{k}\right)\! \subset \mathcal{C}_{\infty}\left(\mathbb{R}, \mathbb{R}^{k}\right) \subset \mathcal{L}_{1}^{\text{loc}}\left(\mathbb{R}, \mathbb{R}^{k}\right)$.

We consider behaviors (systems) defined as the set of \emph{weak} solutions to a linear differential equation:
\begin{equation}
\hspace*{-0.4cm}\mathcal{B} {=} \lbrace \mathbf{w} \in \mathcal{L}_{1}^{\text{loc}}\left(\mathbb{R}, \mathbb{R}^{k}\right) \mid R(\tfrac{d}{dt})\mathbf{w} {=} 0\rbrace, \hspace{0.05cm} R \in \mathbb{R}^{l \times k}[\xi]. \label{eq:bd}
\end{equation}
Here, if $R(\xi) = R_{0} + R_{1}\xi + \ldots + R_{L}\xi^{L}$ and $\mathbf{w} \in \mathcal{C}_{\infty}\left(\mathbb{R}, \mathbb{R}^{k}\right)$, then $R(\tfrac{d}{dt})\mathbf{w} = R_{0}\mathbf{w} + R_{1}\tfrac{d\mathbf{w}}{dt} + \ldots + R_{L}\tfrac{d^{L}\mathbf{w}}{dt^{L}}$ \citep[see][Definition 2.3.7 for the meaning of a weak solution to $R(\tfrac{d}{dt})\mathbf{w} = 0$ when $\mathbf{w}$ is not necessarily differentiable]{JWIMTSC}. Particular attention is paid to the special class of state-space systems:
\begin{align}
&\hspace*{-0.1cm} \mathcal{B}_{s} {=} \lbrace (\mathbf{u}, \mathbf{y}, \mathbf{x}) \in \mathcal{L}_{1}^{\text{loc}}\left(\mathbb{R}, \mathbb{R}^{n}\right) {\times} \mathcal{L}_{1}^{\text{loc}}\left(\mathbb{R}, \mathbb{R}^{n}\right) {\times} \mathcal{L}_{1}^{\text{loc}}\left(\mathbb{R}, \mathbb{R}^{d}\right) \nonumber\\
& \hspace{1.0cm} \text{such that } \tfrac{d\mathbf{x}}{dt} = A\mathbf{x} + B\mathbf{u} \text{ and } \mathbf{y} = C\mathbf{x} + D\mathbf{u}\rbrace, \nonumber \\
&\hspace*{-0.1cm} \text{with } A \in \mathbb{R}^{d \times d}, B \in \mathbb{R}^{d \times n}, C \in \mathbb{R}^{n \times d}, D \in \mathbb{R}^{n \times n}. \label{eq:bne1}
\end{align}
Several properties of state-space systems are listed in Appendix \ref{sec:sot}. In particular, from note \ref{nl:pis3}, if $(\mathbf{u},\mathbf{y},\mathbf{x}) \in \mathcal{B}_{s}$, then $\mathbf{x}$ satisfies the variation of the constants formula almost everywhere, which determines the value $\mathbf{x}(t_{1})$ of $\mathbf{x}$ at an instant $t_{1} \in \mathbb{R}$. Finally, we also consider behaviors obtained by permuting and/or eliminating variables in a behavior $\mathcal{B}$ as in (\ref{eq:bd}). For example, associated with the state-space system $\mathcal{B}_{s}$ in (\ref{eq:bne1}) is the corresponding external behavior $\mathcal{B}_{s}^{(\mathbf{u}, \mathbf{y})} = \lbrace (\mathbf{u},\mathbf{y}) \mid \exists \mathbf{x} \text{ with } (\mathbf{u},\mathbf{y},\mathbf{x}) \in \mathcal{B}_{s}\rbrace$. More generally, for any given $T_{1} \in \mathbb{R}^{l_{1} \times k}, \ldots , T_{n} \in \mathbb{R}^{l_{n} \times k}$ such that $\text{col}(T_{1} \hspace{0.15cm} \cdots \hspace{0.15cm} T_{n}) \in \mathbb{R}^{k \times k}$ is a permutation matrix, and integer $1 \leq m \leq n$, we denote the projection of $\mathcal{B}$ onto $T_{1}\mathbf{w}, \ldots , T_{m}\mathbf{w}$ by
\begin{multline*}
\hspace*{-0.4cm} \mathcal{B}^{(T_{1}\mathbf{w}, \ldots , T_{m}\mathbf{w})} = \lbrace (T_{1}\mathbf{w}, \ldots , T_{m}\mathbf{w}) \mid \exists (T_{m+1}\mathbf{w}, \ldots , T_{n}\mathbf{w}) \\ \text{such that } \mathbf{w} \in \mathcal{B}\rbrace.
\end{multline*}

\section{The positive-real lemma}
\label{sec:pbaprl}

The central role of passivity in systems and control is exemplified by the positive-real lemma (see Lemma \ref{thm:prlos}). The name positive-real (PR) describes a function $G \in \mathbb{R}^{n \times n}(\xi)$ with the properties: (i) $G$ is analytic in $\mathbb{C}_{+}$; and (ii) $G(\bar{\lambda})^{T} + G(\lambda) \geq 0$ for all $\lambda \in \mathbb{C}_{+}$ \citep[see][Theorem 2.7.2 for a well known equivalent condition]{AndVong}. The positive-real lemma then considers a state-space system as in (\ref{eq:bne1}) and provides necessary and sufficient conditions for the transfer function $G(\xi)  = D + C(\xi I {-} A)^{-1}B$ to be PR. Notably, it is assumed that $(A, B)$ is controllable and $(C, A)$ is observable (see notes \ref{nl:pis5} and \ref{nl:pis8}).
\begin{lem}[Positive-real lemma]
\label{thm:prlos}
Let $\mathcal{B}_{s}$ be as in (\ref{eq:bne1}) and let $(A, B)$ be controllable and $(C, A)$ observable. Then the following are equivalent:\newcounter{munit}
\begin{enumerate}[label=\arabic*., ref=\arabic*]
\item Given any $\mathbf{x}_{0} \in \mathbb{R}^{d}$, there exists $S_{a}(\mathbf{x}_{0}) \in \mathbb{R}$ with
$\displaystyle S_{a}(\mathbf{x}_{0}) \coloneqq \sup_{\scriptsize{\begin{smallmatrix}t_{1} \geq t_{0} \in \mathbb{R}, \hspace{0.1cm} (\mathbf{u}, \mathbf{y}, \mathbf{x}) \in \mathcal{B}_{s} \\ \text{with } \mathbf{x}(t_{0}) =  \mathbf{x}_{0}\end{smallmatrix}}}\!\left(\!{-}\int_{t_{0}}^{t_{1}}{ \mathbf{u}^{T}(t)\mathbf{y}(t) dt}\!\right).$ \label{nl:prlc1a}
\item $\displaystyle \sup_{\scriptsize{\begin{smallmatrix}t_{1} \geq t_{0} \in \mathbb{R}, \hspace{0.1cm} (\mathbf{u}, \mathbf{y}, \mathbf{x}) \in \mathcal{B}_{s} \\ \text{with } \mathbf{x}(t_{0}) = 0\end{smallmatrix}}}\!\left(\!-\int_{t_{0}}^{t_{1}}{\mathbf{u}^{T}(t) \mathbf{y}(t) dt}\!\right)\! = 0$ \label{nl:prlc1}.
\item There exist real matrices $X, L_{X}, W_{X}$ such that $X > 0$, $-A^{T}X - XA = L_{X}^{T}L_{X}, C-B^{T}X = W_{X}^{T}L_{X}$, and $D + D^{T} = W_{X}^{T}W_{X}$. \label{nl:prlc2}
\item $G(\xi) \coloneqq D + C(\xi I{-}A)^{-1}B$ is PR. \label{nl:prlc5}
\end{enumerate}
If, in addition, $D+D^{T} > 0$, then the above conditions are equivalent to:
\begin{enumerate}[label=\arabic*., ref=\arabic*]
\setcounter{enumi}{4}
\item There exists a real $X > 0$ such that  $\Pi(X)\coloneqq-A^{T}X-XA - (C^{T}-XB)(D+D^{T})^{-1}(C-B^{T}X) = 0$ and $\text{spec}(A+B(D+D^{T})^{-1}(B^{T}X - C)) \in \overbar{\mathbb{C}}_{-}$.\label{nl:prlc6}
\end{enumerate}
\end{lem}

For a proof of the positive-real lemma, we refer to \citet{JWDSP2, AndVong}. These references also describe links with spectral factorization, which is the concern of the following well known result \citep[Theorem 2]{Youla_SF}:  
\begin{lem}[Youla's spectral factorisation result]
\label{lem:ysfs}
Let $H \in \mathbb{R}^{n \times n}(\xi)$ be para-Hermitian; let $H(j\omega) \geq 0$ for all $\omega \in \mathbb{R}$, $\omega$ not a pole of $H$; and let $\text{normalrank}(H)=r$. There exists a $Z \in \mathbb{R}^{r \times n}(\xi)$ such that (i) $H = Z^{\star}Z$; (ii) $Z$ is analytic in $\mathbb{C}_{+}$; and (iii) $Z(\lambda)$ has full row rank for all $\lambda \in \mathbb{C}_{+}$. Moreover, if $H \in \mathbb{R}^{n \times n}[\xi]$, then $Z \in \mathbb{R}^{r \times n}[\xi]$; if $H(j\omega)$ is analytic for all $\omega \in \mathbb{R}$, then $Z$ is analytic in $\overbar{\mathbb{C}}_{+}$; and if $Z_{1} \in \mathbb{R}^{r \times n}(\xi)$ also satisfies (i)--(iii), then there exists a $T \in \mathbb{R}^{r \times r}$ such that $Z_{1} = TZ$ and $T^{T}T = I$. We call any $Z \in \mathbb{R}^{r \times n}(\xi)$ that satisfies (i)--(iii) a \emph{spectral factor} of $H$.
\end{lem}
\begin{rem}
\label{rem:sfl}
\textnormal{When $G$ is as in Lemma \ref{thm:prlos} with $D+D^{T} > 0$, there exists $W_{X} \in \mathbb{R}^{n \times n}$ with $D+D^{T} = W_{X}^{T}W_{X}$. Then, with $X$ as in condition \ref{nl:prlc6} of Lemma \ref{thm:prlos}, it can be shown that $Z_{X}(\xi) \coloneqq W_{X} +  (W_{X}^{T})^{-1}(C-B^{T}X)(\xi I {-} A)^{-1}B$ is a spectral factor of $G + G^{\star}$ \citep[see][]{JWDSP2}.}
\end{rem}
The assumptions in Lemma \ref{thm:prlos} can be relaxed in three particularly notable ways. First, from \citep[Theorems 1, 3, 8]{JWLSSOC}, conditions \ref{nl:prlc1a}--\ref{nl:prlc5} of Lemma \ref{thm:prlos} are equivalent even if $(C,A)$ is not observable, but $X$ may then be singular in condition \ref{nl:prlc2}. Second, the following are equivalent irrespective of whether $(A,B)$ is controllable or $(C,A)$ is observable: (i) $\text{spec}(A) \in \mathbb{C}_{-}$ and $G(-j  \omega)^{T} + G(j \omega) > 0$ for all $\omega \in \mathbb{R} \cup \infty$; and (ii) the existence of a real symmetric $X \geq 0$ such that $\Pi(X) = 0$ and $\text{spec}(A+B(D+D^{T})^{-1}(B^{T}X - C)) \in \mathbb{C}_{-}$ \citep[Corollary 13.27]{ZDG}. Third, if $\text{spec}(A) \in \mathbb{C}_{-}$, then condition \ref{nl:prlc2} in Lemma \ref{thm:prlos} is equivalent to condition \ref{nl:prlc5} together with the additional condition \citep[equation (4)]{POPRL} (this condition will be discussed in Remark \ref{rem:pcd}).

Nevertheless, the results in these references, and other similar results in the literature \citep[e.g.,][]{CLJKYPL, KPRLNMR}, do not cover several important systems. In particular, they do not consider systems whose transfer functions possess imaginary axis poles. We consider one such system in Example \ref{ex:prlp}. Other important examples include conservative systems, whose transfer functions are lossless PR \citep[see][Chapter 2]{AndVong}.
\begin{example}
\label{ex:prlp}
\textnormal{
Let $\mathcal{B}_{s}$ be as in (\ref{eq:bne1}) with
\begin{equation*}
A = \left[\! \begin{smallmatrix}0& 0& 1\\0& 0& 1\\0 & -1 & 0\end{smallmatrix} \!\right]\!, \hspace{0.1cm} B = \!\left[\! \begin{smallmatrix}1 \\ 0 \\ 0\end{smallmatrix} \!\right]\!, \hspace{0.1cm} C = \!\left[\! \begin{smallmatrix}1& 1& 0\end{smallmatrix}\!\right]\!, \text{ and } D = 1.
\end{equation*}
Here, $(A,B)$ is not controllable. We now show that conditions \ref{nl:prlc1} and \ref{nl:prlc5} of Lemma \ref{thm:prlos} hold for this example, yet condition \ref{nl:prlc1a} does not. First, direct calculation verifies that $G(\xi) = 1 + 1/\xi$, and so condition \ref{nl:prlc5} is satisfied. Second, from the variation of the constants formula (see note \ref{nl:pis3}), $y(t) = x_{1}(t_{0}) + (2\cos (t-t_{0}) - 1)x_{2}(t_{0}) +2\sin (t-t_{0}) x_{3}(t_{0})+ u(t) + \smallint_{t_{0}}^{t}u(\tau)d\tau$ for all $t \geq t_{0}$. Hence, if $\mathbf{x}(t_{0}) = 0$ and $t_{1} \geq t_{0}$, then $\smallint_{t_{0}}^{t_{1}}u(t)y(t)dt = \smallint_{t_{0}}^{t_{1}}u^{2}(t)dt + \tfrac{1}{2}(\smallint_{t_{0}}^{t_{1}}u(\tau) d\tau)^{2} \geq 0$, and so condition \ref{nl:prlc1} is satisfied. Third, with $x_{1}(0) = x_{2}(0) = 0$, $x_{3}(0) = -1$, and $u(t) = \sin(t)$ for all $t \geq 0$, then $y(t) = -\sin(t) - \cos(t)$ for all $t \geq 0$. Thus, for any given positive integer $n$, $-\smallint_{0}^{n\pi}y(t)u(t) dt = \smallint_{0}^{n\pi}\sin^{2}(t)dt + \smallint_{0}^{n\pi}\sin(t)\cos(t)dt = \tfrac{1}{2}n\pi$. It follows that condition \ref{nl:prlc1a} does not hold. Furthermore, it will follow from Theorem \ref{thm:pbt} of this paper that condition  \ref{nl:prlc2} of Lemma \ref{thm:prlos} does not hold for this system.
}
\end{example}

One of the main contributions of this paper is a generalization of the positive-real lemma to include state-space systems that are not necessarily controllable or observable (Theorem \ref{thm:pbt}). In contrast to other papers on this subject, we do not introduce any superfluous assumptions. However, as we will argue in the next section, a state-space system is not a natural starting point for the study of passive systems. Thus, a second major contribution of this paper is a necessary and sufficient condition for the passivity of a general linear system, described by a set of high order differential equations (Theorem \ref{thm:pbtp2}).

\section{Passivity}
\label{sec:pass}
The concept of passivity is relevant to systems whose variables can be partitioned into two sets $\mathbf{i} \in \mathcal{L}_{1}^{\text{loc}}\left(\mathbb{R}, \mathbb{R}^{n}\right)$ and $\mathbf{v} \in \mathcal{L}_{1}^{\text{loc}}\left(\mathbb{R}, \mathbb{R}^{n}\right)$ with the property that  $-\smallint_{t_{0}}^{t_{1}}\mathbf{i}^{T}(t)\mathbf{v}(t)dt$ is the net energy extracted from the system in the interval from $t_{0}$ to $t_{1}$. Passivity has its origins in the study of electric RLCTG networks, for which $\mathbf{i}$ represents the driving-point currents and $\mathbf{v}$ the corresponding driving-point voltages. As shown in \citet{HUGPECBA}, for any given RLCTG network, the driving-point currents and voltages are related by a linear differential equation of the form:
\begin{align}
\label{eq:topbd}
\hspace*{-0.4cm} \mathcal{B} &= \lbrace (\mathbf{i}, \mathbf{v}) \in \mathcal{L}_{1}^{\text{loc}}\left(\mathbb{R}, \mathbb{R}^{n}\right) \times \mathcal{L}_{1}^{\text{loc}}\left(\mathbb{R}, \mathbb{R}^{n}\right) \mid \nonumber \\ &\hspace{0.55cm} 
P(\tfrac{d}{dt})\mathbf{i} = Q(\tfrac{d}{dt})\mathbf{v}, \text{ for some } P, Q \in \mathbb{R}^{n \times n}[\xi]\rbrace.
\end{align}
Note that $(\mathbf{i}, \mathbf{v})$ need not be an \emph{input-output partition} in the sense of \citet{JWIMTSC}. For example: (i) $Q$ is singular for a transformer;\footnote{The behavior of a transformer with turns-ratio matrix $T \in \mathbb{R}^{n_{1} \times n_{2}}$ is determined by the equations $\mathbf{v}_{1} = T^{T}\mathbf{v}_{2}$, and $\mathbf{i}_{2} = -T\mathbf{i}_{1}$, with $\mathbf{v} = \text{col}(\mathbf{v}_{1} \hspace{0.15cm} \mathbf{v}_{2})$ and $\mathbf{i} = \text{col}(\mathbf{i}_{1} \hspace{0.15cm} \mathbf{i}_{2})$).} and (ii) $Q^{-1}P$ is not proper for an inductor.\footnote{For an inductor with inductance $L$, then $Q^{-1}P(\xi) = L\xi$.} Yet it is common for passivity to be defined for systems described using a state-space or input-output representation. This implies assumptions that (i) $Q$ is nonsingular; and (ii) $Q^{-1}P$ is proper. Accordingly, we provide a new definition of passivity for the general system in (\ref{eq:topbd}) that does not depend on such assumptions. Note that this definition extends naturally to non-linear and time-varying systems.
\begin{defn}[Passive system]
\label{def:pb}
The system $\mathcal{B}$ in (\ref{eq:topbd}) is called \emph{passive} if, for any given $(\mathbf{i}, \mathbf{v}) \in \mathcal{B}$ and $t_{0} \in \mathbb{R}$, there exists a $K \in \mathbb{R}$ (dependent on $(\mathbf{i}, \mathbf{v})$ and $t_{0}$) such that if $(\hat{\mathbf{i}}, \hat{\mathbf{v}}) \in \mathcal{B}$ satisfies $(\hat{\mathbf{i}}(t), \hat{\mathbf{v}}(t)) = (\mathbf{i}(t), \mathbf{v}(t))$ for all $t < t_{0}$, then $-\smallint_{t_{0}}^{t_{1}} \hat{\mathbf{i}}^{T}(t)\hat{\mathbf{v}}(t) dt < K$ for all $t_{1} \geq t_{0}$.
\end{defn}
In words, a system is passive if there is an upper bound to the net energy that can be extracted from the system from $t_{0}$ onwards. The upper bound depends on the past of the trajectory, but, given this past, the same upper bound applies to all possible future trajectories.

A detailed discussion of the issues with existing definitions of passivity (and dissipativity) was provided in \citep[Section 8]{JWDDS}. However, for reasons detailed at the end of this section, our definition differs from a similar definition proposed by \citet{JWDDS}. First, we compare Definition \ref{def:pb} to the conditions of the positive-real lemma. Note that it is not essential to follow the discussion in the remainder of this section to understand the main results in the paper.

Condition \ref{nl:prlc1} of Lemma \ref{thm:prlos} is sometimes stated as the definition of passivity for the system in (\ref{eq:bne1}) \citep[e.g.,][Section 2.3]{AndVong}. However, the system in Example \ref{ex:prlp} satisfies this condition but is not passive in the sense of Definition \ref{def:pb}. In other papers, condition \ref{nl:prlc1a} of Lemma \ref{thm:prlos} is stated as the definition for passivity \citep[e.g.,][]{JWDSP2}. It is shown in \citet{THVA} that this is consistent with Definition \ref{def:pb} when considering systems with a state-space realization as in (\ref{eq:bne1}), where $\mathbf{i} = \mathbf{u}$ and $\mathbf{v} = \mathbf{y}$. However, as mentioned earlier, there are systems that are passive in the sense of Definition \ref{def:pb} that cannot be represented in this form. Specifically, as will be shown in Lemma \ref{lem:sstkd}, condition \ref{nl:prlc1a} of Lemma \ref{thm:prlos} only applies to systems of the form:
\begin{multline}
\label{eq:topbdpr}
\hspace*{-0.3cm} \tilde{\mathcal{B}} = \lbrace (\mathbf{u}, \mathbf{y}) \in \mathcal{L}_{1}^{\text{loc}}\left(\mathbb{R}, \mathbb{R}^{n}\right) \times \mathcal{L}_{1}^{\text{loc}}\left(\mathbb{R}, \mathbb{R}^{n}\right) \mid \\ \hspace{-0.5cm} \tilde{P}(\tfrac{d}{dt})\mathbf{u} = \tilde{Q}(\tfrac{d}{dt})\mathbf{y}, \text{ where } \tilde{P}, \tilde{Q} \in \mathbb{R}^{n \times n}[\xi], \\ \hspace{1.2cm} \tilde{Q} \text{ is nonsingular, and } \tilde{Q}^{-1}\tilde{P} \text{ is proper}\rbrace.
\end{multline}
Thus, this condition does not cover systems of the form of (\ref{def:pb}) for which either $Q$ is singular or $Q^{-1}P$ is not proper.

Definition \ref{def:pb} is similar to a definition for dissipativity proposed in \citep[Section 8]{JWDDS} and used by \citet{HugNa} (note that it is straightforward to generalize Definition \ref{def:pb} to the framework of dissipative systems). In \citet{HugNa}, the system $\mathcal{B}$ in (\ref{eq:topbd}) was called passive if, given any $(\mathbf{i}, \mathbf{v}) \in \mathcal{B}$ and any $t_{0} \in \mathbb{R}$, there exists a $K \in \mathbb{R}$ (dependent on $(\mathbf{i}, \mathbf{v})$ and $t_{0}$) such that $-\smallint_{t_{0}}^{t_{1}} \mathbf{i}^{T}(t)\mathbf{v}(t) dt < K$ for all $t_{1} \geq t_{0}$. Evidently, if $\mathcal{B}$ in (\ref{eq:topbd}) is passive in the sense of Definition \ref{def:pb}, then $\mathcal{B}$ is also passive in the sense of \citet{JWDDS, HugNa}. It can also be shown that the converse is true.\footnote{Minor adjustments can be made to the proof given in this paper to show that if $\mathcal{B}$ is passive in the sense of \citet{HugNa}, then condition \ref{nl:pbtc2} of Theorem \ref{thm:pbt} holds.} However, Definition \ref{def:pb} is a more accurate statement of the physical property of passivity (when extended to time-varying and non-linear systems), as the following example demonstrates.
\begin{example}
\textnormal{
Consider the behavior $\mathcal{B} = \lbrace(u, y) \in \mathcal{L}_{1}^{\text{loc}}\left(\mathbb{R}, \mathbb{R}\right) \times \mathcal{L}_{1}^{\text{loc}}\left(\mathbb{R}, \mathbb{R}\right) \mid \exists x \in \mathcal{L}_{1}^{\text{loc}}\left(\mathbb{R}, \mathbb{R}\right)$ with (i) $x(t) = 0$ and $y(t) = 0$ for all $t < 0$; (ii) $\tfrac{dx}{dt}(t) = u(t)$ and $y(t) = 0$ for all $0\leq t<1$; (iii) $\tfrac{dx}{dt}(t) = u(t)$ and $y(t) = 2x(t)$ for all $1\leq t<2$; and (iv) $\tfrac{dx}{dt}(t) = 0$ and $y(t) = 0$ for all $t\geq 2\rbrace$. Thus, if either $t_{0} \geq 2$ or $t_{1} \leq 1$, then $-\smallint_{t_{0}}^{t_{1}} u(t)y(t)dt = 0$; and if instead $t_{1} > t_{0}$, $t_{0} < 2$, and $t_{1} > 1$, then $-\smallint_{t_{0}}^{t_{1}} u(t)y(t)dt = -[x^{2}(t)]_{\max (1,t_{0})}^{\min (2,t_{1})} = -[(\smallint_{0}^{t}u(\tau)d\tau)^{2}]_{\max (1,t_{0})}^{\min (2,t_{1})} \leq (\smallint_{0}^{\max(1, t_{0})}u(\tau)d\tau)^{2}$. It follows that, given any $t_{0} \in \mathbb{R}$, there exists a $K \in \mathbb{R}$ depending on $t_{0}$ and $(u, y)$ such that $-\smallint_{t_{0}}^{t_{1}} u(t)y(t)dt < K$ for all $t_{1} \geq t_{0}$, and so $\mathcal{B}$ is passive in the sense of \citet{HugNa}. On the other hand, for any given $(u, y) \in \mathcal{B}$, $t_{0} < 1$, and $K > 0$, there exists $(\hat{u}, \hat{y}) \in \mathcal{B}$ with $(\hat{u}(t), \hat{y}(t)) = (u(t), y(t))$ for all $t < t_{0}$ such that $-\smallint_{t_{0}}^{t_{1}}\hat{u}(t)\hat{y}(t)dt \geq K$ (e.g., let $\hat{u}(t) = (\sqrt{K} - \smallint_{0}^{t_{0}}u(\tau)d\tau)/(1{-}t_{0})$ for all $t_{0} \leq t < 1$, and $\hat{u}(t) = -\sqrt{K}$ for all $t \geq 1$). Thus, if $t_{0} < 1$, then an arbitrarily large amount of energy can be extracted from this system from $t_{0}$ onwards, and this system is not passive in the sense of Definition \ref{def:pb}.
}
\end{example}

Motivated by electric (RLCTG) networks, we have introduced a definition for passivity for the system in (\ref{eq:topbd}). The classical theory of electric networks provides necessary and sufficient conditions on $P$ and $Q$ for the system in (\ref{eq:topbd}) to be realized by an RLCTG network \emph{providing $P$ and $Q$ are left coprime}. Yet, as emphasised in \citet{camwb}, such conditions are unknown in cases when $P$ and $Q$ are not left coprime. More fundamentally, in these cases, necessary and sufficient conditions on $P$ and $Q$ for the system in (\ref{eq:topbd}) to be \emph{passive} are also unknown. Such conditions are provided in Theorem \ref{thm:pbtp2} of this paper.

\section{The passive behavior theorem}
\label{sec:pbts}

In this section, we present our new passive behavior theorem in two parts. The theorems use our new concept of a \emph{positive-real pair}, which we define as follows:
\begin{defn}
\label{def:prp}
Let $P, Q \in \mathbb{R}^{n \times n}[\xi]$. 
We call $(P, Q)$ a \emph{positive-real pair} if the following conditions hold:
\begin{enumerate}[label=\arabic*., ref=\arabic*]
\item $P(\lambda)Q(\bar{\lambda})^{T} + Q(\lambda)P(\bar{\lambda})^{T} \geq 0$ for all $\lambda \in \overbar{\mathbb{C}}_{+}$. \label{nl:prpc1}
\item $\text{rank}([P \hspace{0.25cm} {-}Q](\lambda)) = n$ for all $\lambda \in \overbar{\mathbb{C}}_{+}$. \label{nl:prpc2} 
\item If $\mathbf{p} \in \mathbb{R}^{n}[\xi]$ and $\lambda \in \mathbb{C}$ satisfy $\mathbf{p}^{T}(PQ^{\star} + QP^{\star}) = 0$ and $\mathbf{p}(\lambda)^{T}[P \hspace{0.25cm} {-}Q](\lambda) = 0$, then $\mathbf{p}(\lambda) = 0$. \label{nl:prpc3}
\end{enumerate}
\end{defn}
\begin{rem}\textnormal{A key result in behavioral theory is that any behavior $\mathcal{B}$ as in (\ref{eq:bd}) has a \emph{controllable} part ($\mathcal{B}_{c}$ in Lemma \ref{lem:dcab}) and an \emph{autonomous} part ($\mathcal{B}_{a}$ in Lemma \ref{lem:dcab}). As will be shown in Section \ref{sec:pbprp}, the conditions in Definition \ref{def:prp} can be understood in terms of $\mathcal{B}_{c}$ and $\mathcal{B}_{a}$. Roughly speaking, the passivity of $\mathcal{B}_{c}$ implies condition \ref{nl:prpc1}; the stability of $\mathcal{B}_{a}$ implies condition \ref{nl:prpc2}, as does the stabilizability of $\mathcal{B}$ (see note \ref{nl:bt3}); and condition \ref{nl:prpc3} is a coupling condition between the trajectories in $\mathcal{B}_{a}$ and the so-called \emph{lossless} trajectories in $\mathcal{B}_{c}$. In particular, if the transfer function from $\mathbf{i}$ to $\mathbf{v}$ is lossless PR \citep[see][Chapter 2]{AndVong}, then $PQ^{\star} + QP^{\star} = 0$, and condition \ref{nl:prpc3} implies that $P$ and $Q$ are left coprime, so $\mathcal{B}$ is controllable (see note \ref{nl:bt3}).}
\end{rem}

We note that condition \ref{nl:prpc1} of Definition \ref{def:prp} is a natural generalization of a positive-real transfer function $Q^{-1}P$ to the case with $Q$ singular. Yet, as discussed in Section \ref{sec:pass}, this condition is not sufficient for the behavior $\mathcal{B}$ in (\ref{eq:topbd}) to be passive. As the following theorem demonstrates, conditions \ref{nl:prpc2} and \ref{nl:prpc3} are also required to obtain a necessary and sufficient condition for passivity. 

\begin{thm}[Passive behavior theorem, Part 1]
\label{thm:pbtp2}
Let $\mathcal{B}$ be as in (\ref{eq:topbd}). Then the following are equivalent:
\begin{enumerate}[label=\arabic*., ref=\arabic*, leftmargin=0.5cm]
\item $\mathcal{B}$ is passive. \label{nl:pbt2c1}
\item $(P, Q)$ is a positive-real pair. \label{nl:pbt2c2}
\item There exist compatible partitions $\mathbf{i} = (\mathbf{i}_{1}, \mathbf{i}_{2})$ and $\mathbf{v} = (\mathbf{v}_{1}, \mathbf{v}_{2})$ such that $\tilde{\mathcal{B}} \coloneqq \mathcal{B}^{(\text{col}(\mathbf{i}_{1} \hspace{0.15cm} \mathbf{v}_{2}), \text{col}(\mathbf{v}_{1} \hspace{0.15cm} \mathbf{i}_{2}))}$ takes the form of (\ref{eq:topbdpr}), and $\tilde{\mathcal{B}}$ is passive.\label{nl:pbt2c3}
\end{enumerate}
\end{thm}
\begin{rem}
\textnormal{
It is also the case that the conditions in Theorem \ref{thm:pbtp2} hold if and only if $\mathcal{B}$ is the driving-point behavior of an electric RLCTG network \citep{HUGPECBA}.
}
\end{rem}

\begin{rem}
\textnormal{
In the terminology of behavioral theory, condition \ref{nl:pbt2c3} of Theorem \ref{thm:pbtp2} implies that if $\mathcal{B}$ in (\ref{eq:topbd}) is passive then there exists an input-output partition \emph{with the property that $\mathbf{i}^{T}\mathbf{v} = \mathbf{u}^{T}\mathbf{y}$} (in the context of electric networks, the input $\text{col}(\mathbf{i}_{1} \hspace{0.15cm} \mathbf{v}_{2})$ contains exactly one variable, either current or voltage, for each port of the network). It is well known that, if $\mathcal{B}$ is as in (\ref{eq:topbd}) and $\text{normalrank}([P \hspace{0.25cm} {-}Q]) = n$, then there exists an \emph{input-output partitioning} of $\text{col}(\mathbf{i} \hspace{0.15cm}\mathbf{v})$ into $\mathbf{u} \in \mathcal{L}_{1}^{\text{loc}}\left(\mathbb{R}, \mathbb{R}^{n}\right)$ and $\mathbf{y} \in \mathcal{L}_{1}^{\text{loc}}\left(\mathbb{R}, \mathbb{R}^{n}\right)$, for which $\tilde{\mathcal{B}} \coloneqq \mathcal{B}^{(\mathbf{u}, \mathbf{y})}$ takes the form of (\ref{eq:topbdpr}) \citep[Section 3.3]{JWIMTSC}. However, this does not suffice to show condition \ref{nl:pbt2c3} in Theorem \ref{thm:pbtp2}. For example, for the system
\begin{equation*}
\begin{bmatrix}0& \tfrac{d}{dt}+1\\ 0& 0\end{bmatrix}\begin{bmatrix}i_{1}\\ i_{2}\end{bmatrix} = \begin{bmatrix}0& 0\\ 0& \tfrac{d}{dt}+2\end{bmatrix}\begin{bmatrix}v_{1}\\ v_{2}\end{bmatrix},
\end{equation*} 
it can be shown that there is no input-output partition with the property that $i_{1}v_{1} + i_{2}v_{2} = \mathbf{u}^{T}\mathbf{y}$.
}
\end{rem}

Theorem \ref{thm:pbtp2} allows us to apply the following results from \citet{JW_TS, PRSMLS, THBRSF} on state-space realizations of behaviors.

\begin{lem}
\label{lem:sstkd}
Let $\mathcal{B}_{s}$ be as in (\ref{eq:bne1}). Then there exist polynomial matrices $\tilde{M}, \tilde{N}, \tilde{P}$ and $\tilde{Q}$ such that
\begin{enumerate}[label=\arabic*., ref=\arabic*, leftmargin=0.5cm]
\item $\tilde{M} \in \mathbb{R}^{n \times n}[\xi]$ and $\tilde{N} \in \mathbb{R}^{n \times d}[\xi]$ are left coprime;\label{nl:behrl1}
\item $\tilde{M}(\xi)C = \tilde{N}(\xi)(\xi I - A)$;\label{nl:behrl2}
\item $\tilde{P} \coloneqq \tilde{N}B + \tilde{M}D$ and $\tilde{Q} \coloneqq \tilde{M}$.\label{nl:behrl3}
\end{enumerate}
Furthermore, if $\tilde{M}, \tilde{N}, \tilde{P}$ and $\tilde{Q}$ satisfy conditions \ref{nl:behrl1}--\ref{nl:behrl3}, then $\tilde{\mathcal{B}} \coloneqq \mathcal{B}_{s}^{(\mathbf{u}, \mathbf{y})}$ takes the form of (\ref{eq:topbdpr}).

Now, let $\tilde{\mathcal{B}}$ take the form of (\ref{eq:topbdpr}). Then there exists $\mathcal{B}_{s}$ as in (\ref{eq:bne1}) such that $\tilde{\mathcal{B}} = \mathcal{B}_{s}^{(\mathbf{u}, \mathbf{y})}$. Also, for any such $\mathcal{B}_{s}$, there exist $\tilde{M}$ and $\tilde{N}$ such that conditions \ref{nl:behrl1}--\ref{nl:behrl3} hold.
\end{lem}

In the next theorem, we consider the state-space system $\mathcal{B}_{s}$ in (\ref{eq:bne1}), and we provide necessary and sufficient conditions for $\mathcal{B}_{s}^{(\mathbf{u}, \mathbf{y})}$ to be passive. This generalizes the positive-real lemma (Lemma \ref{thm:prlos}) to state-space systems that \emph{need not be controllable or observable}.

\begin{thm}[Passive behavior theorem, Part 2]
\label{thm:pbt}
Let $\mathcal{B}_{s}$ be as in (\ref{eq:bne1}); let $\tilde{P}$, $\tilde{Q}$ be as in Lemma \ref{lem:sstkd}; and let $G(\xi) \coloneqq D + C(\xi I - A)^{-1}B$. Then the following are equivalent:
\begin{enumerate}[label=\arabic*., ref=\arabic*]
\item $\tilde{\mathcal{B}} \coloneqq \mathcal{B}_{s}^{(\mathbf{u}, \mathbf{y})}$ is passive. \label{nl:pbtc1}
\item $(\tilde{P}, \tilde{Q})$ is a positive-real pair.\label{nl:pbtc2}
\item There exist real matrices $X, L_{X}, W_{X}$ such that $X \geq 0$, $-A^{T}X - XA = L_{X}^{T}L_{X}, C-B^{T}X = W_{X}^{T}L_{X}$, and $D + D^{T} = W_{X}^{T}W_{X}$.\label{nl:pbtc3}
\item There exist real matrices $X, L_{X}, W_{X}$ as in condition \ref{nl:pbtc3} that have the additional property that $W_{X} + L_{X}(\xi I {-} A)^{-1}B$ is a spectral factor of $G+ G^{\star}$. \label{nl:pbtc4}
\end{enumerate}

If, in addition, $D+D^{T} > 0$, then the above conditions are equivalent to:
\begin{enumerate}[label=\arabic*., ref=\arabic*]
\setcounter{enumi}{4}
\item There exists a real $X \geq 0$ such that $\Pi(X) \coloneqq {-}A^{T}X{-}XA {-} (C^{T}{-}XB)(D{+}D^{T})^{-1}(C{-}B^{T}X) {=} 0$.\label{nl:pbtc5}
\end{enumerate}

Now, suppose conditions \ref{nl:pbtc1}--\ref{nl:pbtc4} hold. Then:
\begin{enumerate}[label=(\roman*), ref=(\roman*)]
\item  If $(C, A)$ is observable and $X$ is as in condition \ref{nl:pbtc3}, then (a) $X > 0$; and (b) $\text{spec}(A) \in \overbar{\mathbb{C}}_{-}$.\label{nl:pbt2ac1}
\item If $D + D^{T} > 0$ and $X$ is as in condition \ref{nl:pbtc4}, then (a) $\Pi(X) = 0$; and (b) $\text{spec}(A+B(D+D^{T})^{-1}(B^{T}X - C)) \in \overbar{\mathbb{C}}_{-}$ if and only if $\text{spec}(A) \in \overbar{\mathbb{C}}_{-}$.\label{nl:pbt2ac2}
\end{enumerate}
\end{thm}

\begin{rem}
\textnormal{
Note that, if the conditions in Theorem \ref{thm:pbt} hold for one state-space realization $\mathcal{B}_{s}$ of $\tilde{\mathcal{B}} \coloneqq \mathcal{B}_{s}^{(\mathbf{u}, \mathbf{y})}$, then they hold for \emph{all} state-space realizations of $\tilde{\mathcal{B}}$. Note also that $\tilde{P}$ and $\tilde{Q}$ are not uniquely defined in that theorem, but it is straightforward to show that condition \ref{nl:pbtc2} is invariant of the specific choice of matrices.}
\end{rem}

\begin{rem}
\label{rem:sfd}
\textnormal{Let $X, L_{X}, W_{X}$ be as in condition \ref{nl:pbtc3} of Theorem \ref{thm:pbt}, let $(\mathbf{u}, \mathbf{y}, \mathbf{x}) \in \mathcal{B}_{s}$, and let $t_{0} \leq t_{1} \in \mathbb{R}$. Since $\mathbf{x}$ is absolutely continuous, then integration by parts gives}
\begin{multline*}
\int_{t_{0}}^{t_{1}}\mathbf{u}^{T}(t)\mathbf{y}(t) + \mathbf{y}^{T}(t)\mathbf{u}(t) dt -  \left[\mathbf{x}^{T}(t)X\mathbf{x}(t)\right]_{t_{0}}^{t_{1}} \\ = \int_{t_{0}}^{t_{1}}(L_{X}\mathbf{x} + W_{X}\mathbf{u})^{T}(t)(L_{X}\mathbf{x} + W_{X}\mathbf{u})(t)dt \geq 0.
\end{multline*}
\textnormal{With the notation $S(\mathbf{x}) \coloneqq \tfrac{1}{2}\mathbf{x}^{T}X\mathbf{x}$ for all $\mathbf{x} \in \mathbb{R}^{d}$, it is straightforward to verify that $S$ is a storage function with respect to the supply rate $\mathbf{u}^{T}\mathbf{y}$ in the sense of \citep[Definition 2]{JWDSP1}. It follows from Theorem \ref{thm:pbt} that, if $\tilde{\mathcal{B}} \coloneqq \mathcal{B}_{s}^{(\mathbf{u}, \mathbf{y})}$ is passive (in accordance with the trajectory-based Definition \ref{def:pb}), then $\mathcal{B}_{s}$ has a (non-negative) quadratic state storage function.}
\end{rem}

\begin{rem}
\textnormal{It is instructive to compare Theorems \ref{thm:pbtp2} and \ref{thm:pbt} with papers by \citet{camwb, PBDUS}, which consider cyclo-dissipativity in the behavioral framework. The reader who is unfamiliar with these papers may prefer to skip straight to Section \ref{sec:pbprp}.}

\textnormal{In \citet{camwb, PBDUS}, cyclo-dissipativity is defined using the formalism of quadratic differential forms (see Appendix \ref{sec:bdf}). With $\mathcal{B}$ as in (\ref{eq:topbd}), then $\mathcal{B}_{\mathcal{C}^{\infty}} \coloneqq \mathcal{B} \cap \mathcal{C}^{\infty}\left(\mathbb{R}, \mathbb{R}^{n}\right) \times \mathcal{C}^{\infty}\left(\mathbb{R}, \mathbb{R}^{n}\right)$ is called \emph{cyclo-dissipative} with respect to the supply rate $\mathbf{i}^{T}\mathbf{v}$ (or \emph{cyclo-passive}) if there exists a quadratic differential form $Q_{\psi}$ such that $\mathbf{i}^{T}\mathbf{v} \geq \tfrac{d}{dt}Q_{\psi}(\text{col}(\mathbf{i} \hspace{0.15cm} \mathbf{v}))$ for all $(\mathbf{i}, \mathbf{v}) \in \mathcal{B}_{\mathcal{C}^{\infty}}$ \citep[Definition 3.1]{PBDUS}. Also, $\mathcal{B}_{\mathcal{C}^{\infty}}$ is called \emph{strictly cyclo-dissipative} with respect to the supply rate $\mathbf{i}^{T}\mathbf{v}$ (or \emph{strictly cyclo-passive}) if there exists a quadratic differential form $Q_{\psi}$ and an $\epsilon > 0$ such that $\mathbf{i}^{T}\mathbf{v} \geq \tfrac{d}{dt}Q_{\psi}(\text{col}(\mathbf{i} \hspace{0.15cm} \mathbf{v})) + \epsilon (\mathbf{i}^{T}\mathbf{i} + \mathbf{v}^{T}\mathbf{v})$ for all $(\mathbf{i}, \mathbf{v}) \in \mathcal{B}_{\mathcal{C}^{\infty}}$ \citep[Definition 3.2]{PBDUS}. In these definitions, $Q_{\psi}$ is called a \emph{storage function} \citep[Definition 4.2]{TWESFSF}, which is called \emph{non-negative} if  $Q_{\psi}(\text{col}(\mathbf{i} \hspace{0.15cm} \mathbf{v}))(t) \geq 0$ for all $(\mathbf{i}, \mathbf{v}) \in \mathcal{B}_{\mathcal{C}^{\infty}}$ and all $t \in \mathbb{R}$. \citet{camwb} considered cyclo-passive single-input single-output systems, while \citet{PBDUS} considered a class of strictly cyclo-dissipative systems that includes the strictly cyclo-passive systems.\footnote{Note that these papers use the word dissipative for what we call cyclo-dissipative systems. We reserve the word dissipative for systems that have a non-negative storage function, as in \citet{JWDSP1}).}}

\textnormal{It can be shown that there are cyclo-passive systems that are not passive, and there are passive systems that are not strictly cyclo-passive. Thus the problems considered in \citet{camwb, PBDUS} are not equivalent to the problem considered in this paper. It can also be shown from Theorems \ref{thm:pbtp2} and \ref{thm:pbt} and Remark \ref{rem:sfd} that $\mathcal{B}_{\mathcal{C}^{\infty}}$ is passive in accordance with Definition \ref{def:pb} if and only if $\mathcal{B}_{\mathcal{C}^{\infty}}$ is cyclo-passive with a non-negative storage function. However, there are two notable reasons why we have not defined a passive system as a cyclo-passive system with a non-negative storage function. First, as discussed in \citet{JWDDS}, it is preferable to define passivity without invoking an a-priori assumption of the existence of a quadratic storage function. This is one of the main benefits of Definition \ref{def:pb}. Second, we note that there is no consensus on the appropriate definition of a cyclo-dissipative system. This concerns the issue of whether to allow for \emph{unobservable storage functions}, as arise in electric networks \citep[see][]{JW_HVDS}. As shown in that paper, there are systems that are not cyclo-dissipative (with respect to a given supply rate), but do possess an unobservable storage function with respect to that supply rate \citep[Section VI]{JW_HVDS}. This issue does not arise with the definition of passivity given in this paper.
}

\textnormal{We also note that \citet{camwb, PBDUS} invoke assumptions that are not present in this paper. In \citet{camwb}, only single-input single-output systems are considered (i.e., $n = 1$ for $\mathcal{B}$ in (\ref{eq:topbd})), for which condition \ref{nl:prpc3} in Definition \ref{def:prp} takes the much simpler form: if $PQ^{\star} + QP^{\star} = 0$, then $[P \hspace{0.25cm} {-}Q](\lambda)$ has full row rank for all $\lambda \in \mathbb{C}$. Also, \citet{camwb} assume that there are no uncontrollable imaginary axis modes (i.e., $\text{rank}([P \hspace{0.25cm} {-}Q](j\omega))$ is constant for all $\omega \in \mathbb{R}$). In contrast, we prove that this condition must hold if $\mathcal{B}$ is passive (note, however, that there may exist $\omega \in \mathbb{R}$ such that $\det(P(j\omega)) = 0$ and/or $\det(Q(j\omega)) = 0$).}

\textnormal{In \citet{PBDUS}, only strictly cyclo-dissipative systems are considered. If $\mathcal{B}$ in (\ref{eq:topbd}) is strictly cyclo-passive, then it can be shown that 1.\ $Q(\lambda)$ and $P(\lambda)$ are nonsingular for all $\lambda \in \overbar{\mathbb{C}}_{+}$; and 2.\ $P(j\omega)Q(-j\omega)^{T} + Q(j\omega)P(-j\omega)^{T}$ is nonsingular for all $\omega \in \mathbb{R}$. The first condition implies that condition \ref{nl:prpc2} of Definition \ref{def:prp} holds (but the converse implication does not hold). Similarly, the second condition implies that condition \ref{nl:prpc3} of Definition \ref{def:prp} holds (again, the converse implication does not hold). Also, the proof of the main results in \citet{PBDUS} used algebraic Riccati equations and Hamiltonian matrices. This approach cannot be used in this paper as it is possible that $D + D^{T}$ is singular in Theorem \ref{thm:pbt}.
}
\end{rem}

\section{Passive behaviors and positive-real pairs}
\label{sec:pbprp}

In Section \ref{sec:pbaprl}, we showed that the system in Example \ref{ex:prlp} has a positive-real transfer function, yet is not passive. For that system, it can be shown that $\mathcal{B}_{s}^{(\mathbf{u}, \mathbf{y})} \eqqcolon \tilde{\mathcal{B}} = \lbrace (u, y) \in \mathcal{L}_{1}^{\text{loc}}\left(\mathbb{R}, \mathbb{R}\right) \times \mathcal{L}_{1}^{\text{loc}}\left(\mathbb{R}, \mathbb{R}\right) \mid (\tfrac{d^{2}}{dt^{2}}+1)(\tfrac{d}{dt}+1)u = (\tfrac{d^{2}}{dt^{2}}+1)\tfrac{d}{dt}y\rbrace$. In particular, if $\tilde{\mathcal{B}}$ is passive, then $\tilde{\mathcal{B}}_{c} = \lbrace (u, y) \in \mathcal{L}_{1}^{\text{loc}}\left(\mathbb{R}, \mathbb{R}\right) \times  \mathcal{L}_{1}^{\text{loc}}\left(\mathbb{R}, \mathbb{R}\right) \mid (\tfrac{d}{dt}+1)u = \tfrac{dy}{dt}\rbrace$ must be passive, and it follows that the transfer function $G(\xi) = 1+1/\xi$ must be PR. But this condition is not sufficient for $\tilde{\mathcal{B}}$ to be passive since there are trajectories in $\tilde{\mathcal{B}}$ with $(\tfrac{d^{2}}{dt^{2}}+1)((\tfrac{d}{dt}+1)u - \tfrac{dy}{dt}) = 0$ but $(\tfrac{d}{dt}+1)u  \not\equiv \tfrac{dy}{dt}$. 

As the preceding example indicates, the transfer function does not always determine the behavior of the system. In contrast, the behavior \emph{is} always determined by the polynomial matrices corresponding to the differential equations governing the system (i.e., by $P$ and $Q$ in  (\ref{eq:topbd})). Thus, passivity will impose requirements on these polynomial matrices. The purpose of this section is to determine these requirements, resulting in Lemma \ref{lem:prln}. We will first prove some alternative requirements in Lemma \ref{lem:irprpc}, which we then show to be equivalent to the conditions in Lemma \ref{lem:prln}. These alternative requirements relate to the following decomposition of the behavior $\mathcal{B}$ in (\ref{eq:topbd}) into \emph{controllable} and \emph{autonomous} parts:
\begin{lem}
\label{lem:dcab}
Let $\mathcal{B}$ in (\ref{eq:topbd}) satisfy $\text{normalrank}([P \hspace{0.25cm} {-}Q]) = n$. Then there exist $F, \tilde{P}, \tilde{Q}, M, N, U, V, X, Y {\in} \mathbb{R}^{n \times n}[\xi]$ such that
\begin{align}
&\hspace*{-0.15cm} P = F\tilde{P}, \hspace{0.15cm} Q = F\tilde{Q}, \text{ and} \label{eq:dca1} \\
&\hspace*{-0.35cm} \begin{bmatrix}\tilde{P}& -\tilde{Q}\\ U& V\end{bmatrix}\begin{bmatrix}X& M\\ Y& N\end{bmatrix} {=} \begin{bmatrix}I_{n}& 0\\ 0& I_{n}\end{bmatrix} {=} \begin{bmatrix}X& M\\ Y& N\end{bmatrix}\begin{bmatrix}\tilde{P}& -\tilde{Q}\\ U& V\end{bmatrix}. \label{eq:dca2}
\end{align}
Now, let  $F, \tilde{P}, \tilde{Q}, M, N, U, V, X, Y \in \mathbb{R}^{n \times n}[\xi]$ satisfy (\ref{eq:dca1})--(\ref{eq:dca2}), and let (i) $\mathcal{B}_{a} \coloneqq \lbrace (\mathbf{i}, \mathbf{v}) \in \mathcal{L}_{1}^{\text{loc}}\left(\mathbb{R}, \mathbb{R}^{n}\right) \times \mathcal{L}_{1}^{\text{loc}}\left(\mathbb{R}, \mathbb{R}^{n}\right) \mid P(\tfrac{d}{dt})\mathbf{i} = Q(\tfrac{d}{dt})\mathbf{v} \text{ and } U(\tfrac{d}{dt})\mathbf{i} = -V(\tfrac{d}{dt})\mathbf{v} \rbrace$; (ii) $\mathcal{B}_{c} \coloneqq \lbrace (\mathbf{i}, \mathbf{v}) \in \mathcal{L}_{1}^{\text{loc}}\left(\mathbb{R}, \mathbb{R}^{n}\right) \times \mathcal{L}_{1}^{\text{loc}}\left(\mathbb{R}, \mathbb{R}^{n}\right) \mid \tilde{P}(\tfrac{d}{dt})\mathbf{i} = \tilde{Q}(\tfrac{d}{dt})\mathbf{v} \rbrace$; and (iii) $\hat{\mathcal{B}} := \lbrace(\mathbf{i}, \mathbf{v}, \mathbf{i}_{1}, \mathbf{v}_{1}, \mathbf{i}_{2}, \mathbf{v}_{2}) \mid (\mathbf{i}_{1}, \mathbf{v}_{1}) \in \mathcal{B}_{a}, (\mathbf{i}_{2}, \mathbf{v}_{2}) \in \mathcal{B}_{c}, \mathbf{i} = \mathbf{i}_{1}+\mathbf{i}_{2} \text{ and } \mathbf{v} = \mathbf{v}_{1} + \mathbf{v}_{2}\rbrace$. Then
\begin{align}
&\hspace*{-0.35cm}\mathcal{B}_{c} {\cap} \mathcal{C}^{\infty}\left(\mathbb{R}, \mathbb{R}^{n}\right) {\times} \mathcal{C}^{\infty}\left(\mathbb{R}, \mathbb{R}^{n}\right) {=} \lbrace (\mathbf{i}, \mathbf{v}) \mid \exists \mathbf{w} {\in} \mathcal{C}^{\infty}\left(\mathbb{R}, \mathbb{R}^{n}\right) \nonumber \\
& \hspace{0.5cm} \text{such that } \mathbf{i} = M(\tfrac{d}{dt})\mathbf{w} \text{ and } \mathbf{v} = N(\tfrac{d}{dt})\mathbf{w}\rbrace, \label{eq:bcr2} \\
&\hspace*{-0.35cm}\mathcal{B}_{a} = \lbrace (\mathbf{i}, \mathbf{v}) \mid \exists \mathbf{z} \in \mathcal{C}^{\infty}\left(\mathbb{R}, \mathbb{R}^{n}\right) \text{ with } F(\tfrac{d}{dt})\mathbf{z} = 0, \nonumber \\
&\hspace{0.5cm} \text{such that } \mathbf{i} = X(\tfrac{d}{dt})\mathbf{z} \text{ and } \mathbf{v} = Y(\tfrac{d}{dt})\mathbf{z} \rbrace,\label{eq:bar2} \\
&\hspace*{-0.35cm} \text{and } \mathcal{B} = \hat{\mathcal{B}}^{(\mathbf{i}, \mathbf{v})}.
\end{align}
\end{lem}

\begin{pf}
The decomposition in the first part of the lemma statement is not unique, but one such decomposition is obtained by computing a lower echelon form for $[P \hspace{0.25cm} {-}Q]$ (see note \ref{nl:pm3}). This gives a unimodular $W \in \mathbb{R}^{2n \times 2n}[\xi]$ such that $[F \hspace{0.25cm} 0] = [P \hspace{0.25cm} {-}Q]W$. Then $W^{-1} \eqqcolon \hat{W} \in \mathbb{R}^{2n \times 2n}[\xi]$, and by suitably partitioning $\hat{W}$ (resp., $W$) we obtain the polynomial matrices in the first (resp., second) block matrix in (\ref{eq:dca2}). 

To show the second part of the lemma, we note initially that (\ref{eq:bcr2})--(\ref{eq:bar2}) are easily shown from (\ref{eq:dca1})--(\ref{eq:dca2}) and \citep[Theorem 3.2.15]{JWIMTSC}. Now, consider the compatibly partitioned matrices
\begin{equation*}
Z {:=} \left[\!\begin{smallmatrix}F\tilde{P}& -F\tilde{Q}& 0& 0\\
U& V& 0& 0\\
0& 0& \tilde{P}& -\tilde{Q}\\
I& 0& I& 0\\
0& I& 0& I\end{smallmatrix}\!\right]\!, R {:=} \left[\!\begin{smallmatrix}0& 0\\
0& 0\\
0& 0\\
I& 0\\
0& I\end{smallmatrix}\!\right]\!,
W {:=} \left[\!\begin{smallmatrix}I& 0& F& -F\tilde{P}& F\tilde{Q}\\ 
0& I& 0& 0& 0\\
0& 0& I& 0& 0\\
0& 0& 0& I& 0\\
0& 0& 0& 0& I\end{smallmatrix}\!\right]\!,
\end{equation*}
and note that $\hat{\mathcal{B}}$ is the set of locally integrable solutions to $R(\tfrac{d}{dt})\text{col}(\mathbf{i} \hspace{0.15cm} \mathbf{v}) = Z(\tfrac{d}{dt})\text{col}(\mathbf{i}_{1} \hspace{0.15cm} \mathbf{v}_{1} \hspace{0.15cm} \mathbf{i}_{2} \hspace{0.15cm} \mathbf{v}_{2})$. Next, let $Z_{2} \in \mathbb{R}^{2n \times 2n}[\xi]$ be formed from the last four block rows of $Z$. It is straightforward to verify from (\ref{eq:dca2}) that $Z_{2}$ is unimodular. As $W$ is unimodular, then by pre-multiplying $R$ and $Z$ by $W$ we conclude that $\hat{\mathcal{B}}$ is the set of locally integrable solutions to $P(\tfrac{d}{dt})\mathbf{i} = Q(\tfrac{d}{dt})\mathbf{v}$ and $\text{col}(0 \hspace{0.15cm} 0 \hspace{0.15cm} \mathbf{i} \hspace{0.15cm} \mathbf{v}) = Z_{2}(\tfrac{d}{dt})\text{col}(\mathbf{i}_{1} \hspace{0.15cm} \mathbf{i}_{2} \hspace{0.15cm} \mathbf{v}_{1} \hspace{0.15cm} \mathbf{v}_{2})$ (see note \ref{nl:bt2}). In particular, $(\mathbf{i}, \mathbf{v}) \in \mathcal{B}$, and it remains to show that, for any given $(\mathbf{i}, \mathbf{v}) \in \mathcal{B}$, there exist locally integrable $(\mathbf{i}_{1}, \mathbf{v}_{1}, \mathbf{i}_{2}, \mathbf{v}_{2})$ such that $\text{col}(0 \hspace{0.15cm} 0 \hspace{0.15cm} \mathbf{i} \hspace{0.15cm} \mathbf{v}) = Z_{2}(\tfrac{d}{dt})\text{col}(\mathbf{i}_{1} \hspace{0.15cm} \mathbf{i}_{2} \hspace{0.15cm} \mathbf{v}_{1} \hspace{0.15cm} \mathbf{v}_{2})$. Accordingly, for any given $H \in \mathbb{R}^{m \times n}[\xi]$ with $\text{normalrank}(H) = m$, we let $\Delta(H)$ denote the maximum degree of all determinants composed of $m$ columns of $H$. Then, from \citep[Theorem 2.8]{JWPPE}, it suffices to show that there exists a determinant of degree $\Delta([R \hspace{0.25cm} Z])$ formed from the columns in $Z$ together with some of the columns in $R$. 

It can be shown that $\Delta([R \hspace{0.25cm} Z]) = \Delta([\tilde{P} \hspace{0.25cm} {-}\tilde{Q}]) + \text{deg}(\text{det}(F))$ (this follows since any non-zero determinant formed from columns of $[R \hspace{0.25cm} Z]$ must contain: (i) the $2n$ non-zero columns from the first two block rows of $[R \hspace{0.25cm} Z]$, which form a nonsingular matrix whose determinant is $\det(F)$; and (ii) at least $n$ non-zero columns from the third block row). Next, let $\Delta([\tilde{P} \hspace{0.25cm} {-}\tilde{Q}])$ be the degree of the determinant formed from columns $i_{1}, \ldots , i_{n}$ of $[\tilde{P} \hspace{0.25cm} {-}\tilde{Q}]$. It can be shown that the degree of the determinant formed from columns $i_{1}, \ldots, i_{n}$ and $2n + 1, \ldots, 6n$ of $[R \hspace{0.25cm} Z]$ equals that of the determinant formed from columns $i_{1}, \ldots, i_{n}$ and $2n + 1, \ldots, 6n$ of $W[R \hspace{0.25cm} Z]$, which equals $\Delta([\tilde{P} \hspace{0.25cm} {-}\tilde{Q}]) + \text{deg}(\text{det}(F)) = \Delta([R \hspace{0.25cm} Z])$. \qed
\end{pf}

Equations (\ref{eq:bcr2})--(\ref{eq:bar2}) represent the infinitely-often differentiable part of the behavior $\mathcal{B}$ in terms of the five matrices $M, N, X, Y$ and $F \in \mathbb{R}^{n \times n}[\xi]$. In the next lemma, we provide three conditions on these matrices for $\mathcal{B}$ to be passive. These correspond to the conditions:
\begin{enumerate}[label=\arabic*., ref=\arabic*]
\item $\mathcal{B}_{c}$ is passive.\label{nl:ppic1}
\item $\mathcal{B}_{a}$ is stable. (i.e., $(\mathbf{i}_{a}, \mathbf{v}_{a}) \in \mathcal{B}_{a} \Rightarrow \mathbf{i}_{a}(t) \rightarrow 0$ and $\mathbf{v}_{a}(t) \rightarrow 0$ as $t \rightarrow \infty$).\label{nl:ppic2}
\item If $t_{0} \leq t_{1} \in \mathbb{R}$, $(\mathbf{i}_{a}, \mathbf{v}_{a}) \in \mathcal{B}_{a}$, and $(\mathbf{i}_{l},\mathbf{v}_{l}) \in \mathcal{B}_{c} \cap \mathcal{C}^{\infty}\left(\mathbb{R}, \mathbb{R}^{n}\right) \times \mathcal{C}^{\infty}\left(\mathbb{R}, \mathbb{R}^{n}\right)$ with $\mathbf{i}_{l}(t) = \mathbf{v}_{l}(t) = 0$ for all $t < t_{0}$ and $\smallint_{t_{0}}^{t_{1}}\mathbf{i}_{l}^{T}(t)\mathbf{v}_{l}(t) dt = 0$, then $\smallint_{t_{0}}^{t_{1}}(\mathbf{i}_{a}^{T}(t)\mathbf{v}_{l}(t) + \mathbf{v}_{a}^{T}(t)\mathbf{i}_{l}(t))dt = 0$.\label{nl:ppic3}
\end{enumerate}
Condition \ref{nl:ppic1} is to be expected since $\mathcal{B}_{c} \subseteq \mathcal{B}$. Condition \ref{nl:ppic2} is equivalent to $\mathcal{B}$ being stabilizable.\footnote{In fact, it was established in \citet{HugNa} that any passive behavior is stabilizable. However, as discussed in Section \ref{sec:pbaprl}, the definition of passivity in \citet{HugNa} differs from the definition in this paper.} Condition \ref{nl:ppic3} is a coupling condition between the \emph{lossless trajectory} $(\mathbf{i}_{l}, \mathbf{v}_{l})$ and the \emph{autonomous trajectory} $(\mathbf{i}_{a}, \mathbf{v}_{a})$. In fact, this condition also holds when $\mathcal{B}_{a}$ is replaced by $\mathcal{B} \cap \mathcal{E}_{\mathbb{C}_{-}}\left(\mathbb{R}, \mathbb{R}^{n}\right) \times \mathcal{E}_{\mathbb{C}_{-}}\left(\mathbb{R}, \mathbb{R}^{n}\right)$ (an observation which is used in the proof of Theorem \ref{thm:pbt}), and provides the intuition behind the third condition of the following lemma:

\begin{lem}
\label{lem:irprpc}
Let $\mathcal{B}$ be as in (\ref{eq:topbd}) and let $\mathcal{B}$ be passive. Then $\text{normalrank}([P \hspace{0.25cm} {-}Q]) = n$. Furthermore, with $M, N$ and $F$ as in Lemma \ref{lem:dcab}, then
\begin{enumerate}[label=\arabic*., ref=\arabic*]
\item $M(\bar{\lambda})^{T}N(\lambda) + N(\bar{\lambda})^{T}M(\lambda) \geq 0$ for all $\lambda \in \overbar{\mathbb{C}}_{+}$. \label{nl:prpcb1}
\item $F(\lambda)$ is nonsingular for all $\lambda \in \overbar{\mathbb{C}}_{+}$. \label{nl:prpcb2}
\item If $(\mathbf{i}_{s}, \mathbf{v}_{s}) \in  \mathcal{B} \cap \mathcal{E}_{\mathbb{C}_{-}}\left(\mathbb{R}, \mathbb{R}^{n}\right) \times \mathcal{E}_{\mathbb{C}_{-}}\left(\mathbb{R}, \mathbb{R}^{n}\right)$ and $\mathbf{b} \in \mathbb{R}^{n}[\xi]$ satisfies $\mathbf{b}^{\star}(M^{\star}N + N^{\star}M) = 0$, then $\mathbf{b}^{\star}(\tfrac{d}{dt})(M^{\star}(\tfrac{d}{dt})\mathbf{v}_{s} + N^{\star}(\tfrac{d}{dt})\mathbf{i}_{s}) = 0$. \label{nl:prpcb3}
\end{enumerate}
\end{lem}

\begin{pf}
We first show that $n = \text{rank}([P \hspace{0.25cm} {-}Q](\lambda)) = \text{rank}(F(\lambda)[\tilde{P} \hspace{0.25cm} {-}\tilde{Q}](\lambda))$ for all $\lambda \in \overbar{\mathbb{C}}_{+}$. This implies that $\text{normalrank}([P \hspace{0.25cm} {-}Q]) = n$ and condition \ref{nl:prpcb2} holds. We then show condition \ref{nl:prpcb1}, and finally condition \ref{nl:prpcb3}.

{\bf Proof that $\boldsymbol{\text{rank}([P \hspace{0.25cm} {-}Q](\lambda)) = n}$ for all $\boldsymbol{\lambda \in \overbar{\mathbb{C}}_{+}}$.} \hspace{0.3cm}
Suppose instead that there exists $\lambda \in \overbar{\mathbb{C}}_{+}$ such that $\text{rank}([P \hspace{0.25cm} {-}Q](\lambda)) < n$. Then $\text{rank}(P(\lambda)+Q(\lambda)) < n$, and so there exists $0 \neq \mathbf{z}  \in \mathbb{C}^{n}$ such that $(P(\lambda) + Q(\lambda))\mathbf{z} = 0$. Then, with the notation $\mathbf{v}(t) = \mathbf{z}e^{\lambda t} + \bar{\mathbf{z}}e^{\bar{\lambda}t}$ and $\mathbf{i}(t) = -\mathbf{v}(t)$ for all $t \in \mathbb{R}$, we find that $(\mathbf{i}, \mathbf{v}) \in \mathcal{B}$. Also, for any given $t_{1} \geq t_{0} \in \mathbb{R}$, then $-\smallint_{t_{0}}^{t_{1}}\mathbf{i}^{T}(t)\mathbf{v}(t)dt {=} 2\Re{(\mathbf{z}^{T}\mathbf{z}\smallint_{t_{0}}^{t_{1}}e^{2\lambda t}dt)} + 2(\bar{\mathbf{z}}^{T}\mathbf{z})\smallint_{t_{0}}^{t_{1}}e^{2\Re(\lambda) t}dt$. By considering separately the cases $\Im{(\lambda)} = 0$ and $\Im{(\lambda)} \neq 0$, it can be shown that for any given $K \in \mathbb{R}$ there exists $t_{1} \geq t_{0} \in \mathbb{R}$ such that $-\smallint_{t_{0}}^{t_{1}}\mathbf{i}^{T}(t)\mathbf{v}(t)dt \geq K$, whence $\mathcal{B}$ is not passive. Thus, if $\mathcal{B}$ is passive, then $\text{rank}([P \hspace{0.25cm} {-}Q](\lambda)) = n$ for all $\lambda \in \overbar{\mathbb{C}}_{+}$.

{\bf Proof of condition \ref{nl:prpcb1}.} \hspace{0.3cm}
Consider a fixed but arbitrary $\lambda \in \overbar{\mathbb{C}}_{+}$ and $\mathbf{c} \in \mathbb{C}^{n}$; let $\mathbf{z}(t) = \mathbf{c}e^{\lambda t} + \mathbf{\bar{c}}e^{\bar{\lambda}t}$ for all $t \in \mathbb{R}$; let $\mathbf{i} \coloneqq M(\tfrac{d}{dt})\mathbf{z}$ and $\mathbf{v} \coloneqq N(\tfrac{d}{dt})\mathbf{z}$; let $\Psi(\eta, \xi) \coloneqq M(\eta)^{T}N(\xi) + N(\eta)^{T}M(\xi)$; and let $\alpha \coloneqq \mathbf{c}^{T}\Psi(\lambda, \lambda)\mathbf{c}$ and $\beta \coloneqq \bar{\mathbf{c}}^{T}\Psi\left(\bar{\lambda}, \lambda\right)\mathbf{c}$. Then $(\mathbf{i}, \mathbf{v}) \in \mathcal{B}$ by Lemma \ref{lem:dcab}, and 
\begin{equation}
\hspace*{-0.45cm} \int_{t_{0}}^{t_{1}}\!\mathbf{i}^{T}(t)\mathbf{v}(t)dt = \Re\!\left(\!\alpha \!\int_{t_{0}}^{t_{1}}\!e^{2 \lambda t}dt \!\right)\! {+} \beta \!\int_{t_{0}}^{t_{1}}\!e^{2\Re(\lambda)t}dt.\label{eq:ivcep1}
\end{equation}
We will show that if there exists a $\lambda \in \overbar{\mathbb{C}}_{+}$ and $\mathbf{c} \in \mathbb{C}^{n}$ such that $\beta = \bar{\mathbf{c}}^{T}\Psi(\bar{\lambda}, \lambda)\mathbf{c} < 0$, then for any given $K \in \mathbb{R}$ there exists a $t_{1} \geq t_{0}$ with $-\smallint_{t_{0}}^{t_{1}}\mathbf{i}^{T}(t)\mathbf{v}(t)dt \geq K$. This will prove condition \ref{nl:prpcb1}.

Let $\lambda = \sigma + j\omega$ for some $\sigma, \omega \in \mathbb{R}$ with $\sigma \geq 0$, and consider a fixed but arbitrary $K \in \mathbb{R}$. We consider the cases (i) $\omega = 0$; and (ii) $\omega \neq 0$. In case (i), let $\lambda = \bar{\lambda}$ and $\mathbf{c} = \bar{\mathbf{c}}$, so $\alpha = \beta$. Then, from (\ref{eq:ivcep1}), $\smallint_{t_{0}}^{t_{1}}\mathbf{i}^{T}(t)\mathbf{v}(t)dt = 2 \beta \smallint_{t_{0}}^{t_{1}}e^{2\lambda t}dt$, and $\smallint_{t_{0}}^{t_{1}}e^{2\lambda t}dt = (1/2\lambda)(e^{2\lambda t_{1} - 2\lambda t_{0}})$ if $\Re(\lambda) \neq 0$, and $t_{1}- t_{0}$ otherwise. In case (ii), for any given integer $n$, we let $T(n) \in \mathbb{R}$ satisfy $2\omega T(n) = 2\pi (n + 1/4) - \arg(\alpha /(\sigma + j\omega))$ (note, if $T(n) \geq t_{0}$, then $n \geq \omega t_{0}/ \pi + 1/4$ when $\omega > 0$, and $n \leq \omega t_{0}/ \pi - 3/4$ when $\omega < 0$). Then $\arg(\alpha e^{2\lambda T(n)}/\lambda) = \pi / 2$, so from (\ref{eq:ivcep1}) we find that $\smallint_{t_{0}}^{T(n)}\mathbf{i}^{T}(t)\mathbf{v}(t)dt = (\beta (e^{2\Re{(\lambda)}T(n)}{-}e^{2\Re{(\lambda)}t_{0}})/2\Re{(\lambda)}) - \Re(\alpha e^{2\lambda t_{0}}/2\lambda)$ if $\Re{(\lambda)} \neq 0$, and $\beta (T(n) - t_{0}) - \Re(\alpha e^{2\lambda t_{0}}/2\lambda)$ otherwise. In both cases (i) and (ii), if $\beta < 0$, then by taking $t_{1}$ sufficiently large (and letting $t_{1} = T(n)$ in case (ii)) we obtain $-\smallint_{t_{0}}^{t_{1}}\mathbf{i}^{T}(t)\mathbf{v}(t)dt \geq K$.

{\bf Proof of condition \ref{nl:prpcb3}.} \hspace{0.3cm}
Let $\mathbf{b} \in \mathbb{R}^{n}[\xi]$ satisfy $\mathbf{b}^{\star}(M^{\star}N+N^{\star}M) = 0$, let $t_{0} \leq t_{1} \in \mathbb{R}$, and consider a fixed but arbitrary $(\mathbf{i}_{s}, \mathbf{v}_{s}) \in \mathcal{B} \cap \mathcal{E}_{\mathbb{C}_{-}}\left(\mathbb{R}, \mathbb{R}^{n}\right) \times \mathcal{E}_{\mathbb{C}_{-}}\left(\mathbb{R}, \mathbb{R}^{n}\right)$ and $z \in  \mathcal{C}^{\infty}\left(\mathbb{R}, \mathbb{R}\right)$. Then, with the notation $\mathbf{i} \coloneqq M(\tfrac{d}{dt})\mathbf{b}(\tfrac{d}{dt})z + \mathbf{i}_{s}$ and $\mathbf{v} \coloneqq N(\tfrac{d}{dt})\mathbf{b}(\tfrac{d}{dt})z + \mathbf{v}_{s}$, it follows that $(\mathbf{i}, \mathbf{v}) \in \mathcal{B}$ by Lemma \ref{lem:dcab}. Also, with
\begin{align}
& \hspace*{-0.3cm} J_{1} {\coloneqq} \int_{t_{0}}^{t_{1}}((M(\tfrac{d}{dt})\mathbf{b}(\tfrac{d}{dt})z)^{T}(N(\tfrac{d}{dt})\mathbf{b}(\tfrac{d}{dt})z))(t)dt, \text{ and}\nonumber\\
& \hspace*{-0.3cm} J_{2} {\coloneqq} \int_{t_{0}}^{t_{1}}((M(\tfrac{d}{dt})\mathbf{b}(\tfrac{d}{dt})z)^{T}\mathbf{v}_{s} {+} (N(\tfrac{d}{dt})\mathbf{b}(\tfrac{d}{dt})z)^{T}\mathbf{i}_{s})(t) dt,\nonumber \\ 
& \hspace*{-0.3cm}\text{then } \int_{t_{0}}^{t_{1}}(\mathbf{i}^{T}\mathbf{v})(t)dt = J_{1} {+} J_{2} {+} \int_{t_{0}}^{t_{1}}(\mathbf{i}_{s}^{T}\mathbf{v}_{s})(t)dt. \label{eq:intj13d}
\end{align}
Since $\mathbf{b}^{\star}(M^{\star}N + N^{\star}M) = 0$ then, from note \ref{nl:bdf3},
\begin{align*}
&J_{1} {=} \tfrac{1}{2}\left[L_{\Phi_{M\mathbf{b}}}(z, (N\mathbf{b})(\tfrac{d}{dt})z)(t) {+} L_{\Phi_{N\mathbf{b}}}(z, (M\mathbf{b})(\tfrac{d}{dt})z)(t)\right]_{t_{0}}^{t_{1}}, \\
&\text{and }J_{2} = \int_{t_{0}}^{t_{1}}(z(\mathbf{b}^{\star}(\tfrac{d}{dt})(M^{\star}(\tfrac{d}{dt})\mathbf{v}_{s} + N^{\star}(\tfrac{d}{dt})\mathbf{i}_{s})))(t)dt \\
&\hspace{1.75cm} + \left[L_{\Phi_{M\mathbf{b}}}(z, \mathbf{v}_{s})(t) {+} L_{\Phi_{N\mathbf{b}}}(z, \mathbf{i}_{s})(t)\right]_{t_{0}}^{t_{1}}.
\end{align*}
Now, let $g \coloneqq \mathbf{b}^{\star}(\tfrac{d}{dt})(M^{\star}(\tfrac{d}{dt})\mathbf{v}_{s} + N^{\star}(\tfrac{d}{dt})\mathbf{i}_{s})$; let $\psi \in  \mathcal{C}^{\infty}\left(\mathbb{R}, \mathbb{R}\right)$ and $t_{0} \leq t_{1} \in \mathbb{R}$ satisfy $\psi(t) = 0$ for all $t \leq t_{0}$, and $\tfrac{d^{k}\psi}{dt^{k}}(t_{1}) = 0$ ($k = 0, 1, 2, \ldots$); let $z \coloneqq g\psi$; and let $f \coloneqq g^{2}$. Then, $z,f \in  \mathcal{C}^{\infty}\left(\mathbb{R}, \mathbb{R}\right)$; $f(t) \geq 0$ for all $t \in \mathbb{R}$; $J_{1} = 0$; and $J_{2} = \smallint_{t_{0}}^{t_{1}}(f\psi)(t)dt$. Moreover, since $(\mathbf{i}_{s}, \mathbf{v}_{s}) \in \mathcal{B} \cap \mathcal{E}_{\mathbb{C}_{-}}\left(\mathbb{R}, \mathbb{R}^{n}\right) \times \mathcal{E}_{\mathbb{C}_{-}}\left(\mathbb{R}, \mathbb{R}^{n}\right)$, then it is straightforward to show that $\mathbf{i}_{s}^{T}\mathbf{v}_{s} \in \mathcal{E}_{\mathbb{C}_{-}}\left(\mathbb{R}, \mathbb{R}\right)$, and that there exists an $M \in \mathbb{R}$ such that $\smallint_{t_{0}}^{t}(\mathbf{i}_{s}^{T}\mathbf{v}_{s})(t)dt < M$ for all $t \geq t_{0}$. Thus, from (\ref{eq:intj13d}), there exists an $M \in \mathbb{R}$ such that $-\smallint_{t_{0}}^{t_{1}}(\mathbf{i}^{T}\mathbf{v})(t)dt > -\smallint_{t_{0}}^{t_{1}}(f\psi)(t)dt - M$. Finally, we will show that, for any given $K \in \mathbb{R}$, there exist $\psi$ and $t_{1}$ with the properties outlined above that satisfy $-\smallint_{t_{0}}^{t_{1}}(f\psi)(t)dt > K{+}M$. This proves condition \ref{nl:prpcb3}.

Let $\phi(t) = e^{1/(t^{2}-1)}$ for $-1 < t < 1$ with $\phi(t) = 0$ otherwise. Also, for any given integer $k$, let $g_{k}(t) \coloneqq f(t)\phi(t-1-t_{0}-2k)$ for all $t \in \mathbb{R}$. Note that $g_{k} \in \mathcal{C}^{\infty}\left(\mathbb{R}, \mathbb{R}\right)$ and $\smallint_{t_{0}+2k}^{t_{0} + 2(k+1)}g_{k}(t)dt > 0$ ($k = 0,1,\ldots$). Now, let $N$ be a positive integer with $N > K+M$, and let $\psi(t) = -\sum_{k = 0}^{N-1}\phi(t-1-t_{0}-2k)/(\smallint_{t_{0}+2k}^{t_{0} + 2(k+1)}g_{k}(t)dt)$ for all $t \in \mathbb{R}$. It can be verified that $\psi \in \mathcal{C}^{\infty}\left(\mathbb{R}, \mathbb{R}\right)$; $\psi(t) = 0$ for all $t \leq t_{0}$; $\tfrac{d^{l}\psi}{dt^{l}}(t_{0}+2k) = 0$ for $k,l = 0, 1, 2, \ldots$; and $-\smallint_{t_{0}}^{t_{0}+2N}(f\psi)(t)dt = N > K+M$. \qed
\end{pf}

In the next lemma, we present several equivalent conditions to the third condition in Lemma \ref{lem:irprpc}. This leads to two algebraic tests for this condition (see Remark \ref{rem:eac3prp}), and the main result in this section (see Lemma \ref{lem:prln}).
\begin{lem}
\label{lem:prpc3ec}
Let $\mathcal{B}$ be as in (\ref{eq:topbd}); let $\text{rank}([P \hspace{0.25cm} {-}Q](\lambda)) = n$ for all $\lambda \in \overbar{\mathbb{C}}_{+}$; and let $F, \tilde{P}, \tilde{Q}, M, N, U, V, X, Y$, and $\mathcal{B}_{a}$ be as in Lemma \ref{lem:dcab}. Then the following are equivalent:
\begin{enumerate}[label=\arabic*., ref=\arabic*, leftmargin=0.4cm]
\item Condition \ref{nl:prpcb3} of Lemma \ref{lem:irprpc} holds. \label{nl:prpc3c1}
\item If $(\mathbf{i}_{a}, \mathbf{v}_{a}) \in  \mathcal{B}_{a}$ and $\mathbf{b} \in \mathbb{R}^{n}[\xi]$ satisfies $\mathbf{b}^{\star}(M^{\star}N + N^{\star}M) = 0$, then $\mathbf{b}^{\star}(\tfrac{d}{dt})(M^{\star}(\tfrac{d}{dt})\mathbf{v}_{a} + N^{\star}(\tfrac{d}{dt})\mathbf{i}_{a}) = 0$. \label{nl:prpc3c2}
\item If $\mathbf{b} \in \mathbb{R}^{n}[\xi]$ satisfies $\mathbf{b}^{\star}(M^{\star}N + N^{\star}M) = 0$, then there exists $\mathbf{p} \in \mathbb{R}^{n}[\xi]$ such that $\mathbf{b}^{\star}(M^{\star}Y + N^{\star}X) = \mathbf{p}^{T}F$.\label{nl:prpc3c3}
\item If $\mathbf{c} \in \mathbb{R}^{n}[\xi]$ satisfies $\mathbf{c}^{T}(\tilde{P}\tilde{Q}^{\star} + \tilde{Q}\tilde{P}^{\star}) = 0$, then there exists $\mathbf{p} \in \mathbb{R}^{n}[\xi]$ such that $\mathbf{c}^{T} = \mathbf{p}^{T}F$.\label{nl:prpc3c4}
\item If $\mathbf{p} \in \mathbb{R}^{n}[\xi]$ and $\lambda \in \mathbb{C}$ satisfy $\mathbf{p}^{T}(PQ^{\star} + QP^{\star}) = 0$ and $\mathbf{p}(\lambda)^{T}[P \hspace{0.25cm} {-}Q](\lambda) = 0$, then $\mathbf{p}(\lambda) = 0$.\label{nl:prpc3c5}
\end{enumerate}
\end{lem}

\begin{pf}
\ref{nl:prpc3c1} $\iff$ \ref{nl:prpc3c2}. \hspace{0.3cm} That \ref{nl:prpc3c1} $\Rightarrow$ \ref{nl:prpc3c2} follows since $\text{rank}([P \hspace{0.25cm} {-}Q](\lambda)) = n$ for all $\lambda \in \overbar{\mathbb{C}}_{+}$ implies that $F(\lambda)$ is non-singular for all $\lambda \in \overbar{\mathbb{C}}_{+}$, and so $\mathcal{B}_{a} \subseteq \mathcal{B} \cap \mathcal{E}_{\mathbb{C}_{-}}\left(\mathbb{R}, \mathbb{R}^{n}\right) \times \mathcal{E}_{\mathbb{C}_{-}}\left(\mathbb{R}, \mathbb{R}^{n}\right)$ by Lemma \ref{lem:dcab} and \citep[Section 3.2.2]{JWIMTSC}. Then \ref{nl:prpc3c2} $\Rightarrow$ \ref{nl:prpc3c1} since, by Lemma \ref{lem:dcab}, if $(\mathbf{i}_{s}, \mathbf{v}_{s}) \in \mathcal{B} \cap \mathcal{E}_{\mathbb{C}_{-}}\left(\mathbb{R}, \mathbb{R}^{n}\right) \times \mathcal{E}_{\mathbb{C}_{-}}\left(\mathbb{R}, \mathbb{R}^{n}\right)$, then there exists $(\mathbf{i}_{a}, \mathbf{v}_{a}) \in \mathcal{B}_{a}$ and $\mathbf{z} \in \mathcal{C}_{\infty}\left(\mathbb{R}, \mathbb{R}^{n}\right)$ such that $\mathbf{i}_{s} = M(\tfrac{d}{dt})\mathbf{z} + \mathbf{i}_{a}$ and $\mathbf{v}_{s} = N(\tfrac{d}{dt})\mathbf{z} + \mathbf{v}_{a}$.

\ref{nl:prpc3c2} $\iff$ \ref{nl:prpc3c3}. \hspace{0.3cm} To see that \ref{nl:prpc3c2} $\Rightarrow$ \ref{nl:prpc3c3}, note initially from Lemma \ref{lem:dcab} that condition \ref{nl:prpc3c2} implies that if $\mathbf{b} \in \mathbb{R}^{n}[\xi]$ satisfies $\mathbf{b}^{\star}(M^{\star}N+N^{\star}M)$, and $\mathbf{z} \in \mathcal{L}_{1}^{\text{loc}}\left(\mathbb{R}, \mathbb{R}^{n}\right)$ satisfies $F(\tfrac{d}{dt})\mathbf{z} = 0$, then $\mathbf{b}^{\star}(\tfrac{d}{dt})(M^{\star}Y+N^{\star}X)(\tfrac{d}{dt})\mathbf{z} = 0$. From note \ref{nl:bt2b}, this implies that there exists $\mathbf{p} \in \mathbb{R}^{n}[\xi]$ such that $\mathbf{b}^{\star}(M^{\star}Y + N^{\star}X) = \mathbf{p}^{T}F$. Similarly, from Lemma \ref{lem:dcab}, it is straightforward to show that \ref{nl:prpc3c3} $\Rightarrow$ \ref{nl:prpc3c2}.

\ref{nl:prpc3c3} $\iff$ \ref{nl:prpc3c4}. \hspace{0.3cm} Note initially from (\ref{eq:dca2}) that
\begin{align}
\hspace*{-0.35cm}\begin{bmatrix}\tilde{P}& \hspace{0.15cm} {-}\tilde{Q}\end{bmatrix}\! \!\begin{bmatrix}\!-\tilde{Q}^{\star}& V^{\star}\\ \tilde{P}^{\star}& U^{\star}\!\end{bmatrix}\! \!\begin{bmatrix}\!Y^{\star}& X^{\star}\\ N^{\star}& M^{\star}\!\end{bmatrix}\! \!\begin{bmatrix}\!X& M\\ Y& N\!\end{bmatrix}\! = \!\begin{bmatrix}\!I_{n}& \hspace{0.15cm} 0\end{bmatrix}.\label{eq:pqxyr1}
\end{align}
To prove that \ref{nl:prpc3c3} $\Rightarrow$ \ref{nl:prpc3c4}, note that if $\mathbf{c} \in \mathbb{R}^{n}[\xi]$ satisfies $\mathbf{c}^{T}(\tilde{P}\tilde{Q}^{\star} + \tilde{Q}\tilde{P}^{\star}) = 0$, then $\mathbf{b}^{\star} \coloneqq \mathbf{c}^{T}(\tilde{P}V^{\star} - \tilde{Q}U^{\star})$ satisfies $\mathbf{b}^{\star}(M^{\star}N+N^{\star}M) = 0$ by (\ref{eq:pqxyr1}). Thus, from condition \ref{nl:prpc3c3}, there exists $\mathbf{p} \in \mathbb{R}^{n}[\xi]$ such that $\mathbf{b}^{\star}(M^{\star}Y + N^{\star}X) = \mathbf{p}^{T}F$. But $\mathbf{b}^{\star}(M^{\star}Y + N^{\star}X) = \mathbf{c}^{T}(\tilde{P}V^{\star} - \tilde{Q}U^{\star})(M^{\star}Y + N^{\star}X)$, and $\mathbf{c}^{T} = \mathbf{c}^{T}(\tilde{P}V^{\star} - \tilde{Q}U^{\star})(M^{\star}Y + N^{\star}X) = \mathbf{p}^{T}F$ by (\ref{eq:pqxyr1}). The proof of \ref{nl:prpc3c4} $\Rightarrow$ \ref{nl:prpc3c3} is similar.

\ref{nl:prpc3c4} $\iff$ \ref{nl:prpc3c5}. \hspace{0.3cm} To see that \ref{nl:prpc3c4} $\Rightarrow$ \ref{nl:prpc3c5}, we let $r \coloneqq \text{normalrank}(\tilde{P}\tilde{Q}^{\star} + \tilde{Q}\tilde{P}^{\star})$, and we let the rows of $V_{1} \in \mathbb{R}^{(n{-}r) \times n}[\xi]$ be a basis for the left syzygy of $\tilde{P}\tilde{Q}^{\star} + \tilde{Q}\tilde{P}^{\star}$ (see note \ref{nl:pm2}). Then condition \ref{nl:prpc3c4} implies that there exists $\hat{V}_{1} \in \mathbb{R}^{(n{-}r) \times n}[\xi]$ such that $V_{1} = \hat{V}_{1}F$. Since $V_{1}(\lambda)$ has full row rank for all $\lambda \in \mathbb{C}$, then so too must $\hat{V}_{1}(\lambda)$. Since, in addition, $PQ^{\star} + QP^{\star} = F(\tilde{P}\tilde{Q}^{\star} + \tilde{Q}\tilde{P}^{\star})F^{\star}$, then $\hat{V}_{1}(PQ^{\star} + QP^{\star}) = V_{1}(\tilde{P}\tilde{Q}^{\star} + \tilde{Q}\tilde{P}^{\star})F^{\star} = 0$ and $\text{normalrank}(PQ^{\star} + QP^{\star}) = \text{normalrank}(\tilde{P}\tilde{Q}^{\star} + \tilde{Q}\tilde{P}^{\star})$, and we conclude that the rows of $\hat{V}_{1}$ are a basis for the left syzygy of $PQ^{\star} {+} QP^{\star}$. It follows that if $\mathbf{p}^{T}(PQ^{\star} {+} QP^{\star}) = 0$, then there exists $\mathbf{g} {\in} \mathbb{R}^{(n{-}r)}[\xi]$ such that $\mathbf{p}^{T} = \mathbf{g}^{T}\hat{V}_{1}$. If, in addition, $\mathbf{p}(\lambda)^{T}[P \hspace{0.25cm} {-}Q](\lambda) = 0$, then $\mathbf{p}(\lambda)^{T}F(\lambda) = 0$ since $\tilde{P}$ and $\tilde{Q}$ are left coprime, whence $\mathbf{g}(\lambda)^{T}\hat{V}_{1}(\lambda)F(\lambda) = \mathbf{g}(\lambda)^{T}V_{1}(\lambda) = 0$. But $V_{1}(\lambda)$ has full row rank for all $\lambda \in \mathbb{C}$, and we conclude that $\mathbf{g}(\lambda) = 0$ and so $\mathbf{p}(\lambda) = 0$. Finally, to show that \ref{nl:prpc3c5} $\Rightarrow$ \ref{nl:prpc3c4}, we let the rows of $\hat{V}_{1} \in \mathbb{R}^{(n{-}r) \times n}[\xi]$ be a basis for the left syzygy of $PQ^{\star} + QP^{\star}$. Then, from condition \ref{nl:prpc3c5}, we conclude that $\mathbf{c}(\lambda)^{T}\hat{V}_{1}(\lambda)F(\lambda) = 0 \Rightarrow \mathbf{c}(\lambda)^{T} = 0$, and it follows that $\hat{V}_{1}(\lambda)F(\lambda)$ has full row rank for all $\lambda \in \mathbb{C}$. In a similar manner to before, it can then be shown that the rows of $\hat{V}_{1}F$ are a basis for the left syzygy of $\tilde{P}\tilde{Q}^{\star} + \tilde{Q}\tilde{P}^{\star}$. Hence, if $\mathbf{c} \in \mathbb{R}^{n}[\xi]$ satisfies $\mathbf{c}^{T}(\tilde{P}\tilde{Q}^{\star} + \tilde{Q}\tilde{P}^{\star}) = 0$, then there exists $\mathbf{g} \in \mathbb{R}^{(n{-}r)}[\xi]$ such that $\mathbf{c}^{T} = \mathbf{g}^{T}\hat{V}_{1}F$, and by letting $\mathbf{p}^{T} := \mathbf{g}^{T}\hat{V}_{1}$ we obtain condition \ref{nl:prpc3c4}. \qed
\end{pf}

\begin{rem}
\label{rem:eac3prp}
\textnormal{Let $\mathcal{B}, F, \tilde{P}, \tilde{Q}, M, N, U, V, X, Y$, and $\mathcal{B}_{a}$ be as in Lemma \ref{lem:prpc3ec} (with $\text{rank}([P \hspace{0.25cm} {-}Q](\lambda)) = n$ for all $\lambda \in \overbar{\mathbb{C}}_{+}$). Lemma \ref{lem:prpc3ec} leads to two tests that can be implemented by a standard symbolic algebra program (using exact arithmetic if the polynomial matrix coefficients are rational). As in the proof of Lemma \ref{lem:prpc3ec}, let $r \coloneqq \text{normalrank}(PQ^{\star} {+} QP^{\star})$, and note that it is easily shown from the proof of that lemma that $r = \text{normalrank}(M^{\star}N{+}N^{\star}M)$. The two tests are as follows.
\begin{enumerate}[label=\arabic*., ref=\arabic*, leftmargin=0.4cm]
\item Using the matrices $M, N, X, Y$ and $F$:
\begin{enumerate}
\item Compute a $V \in \mathbb{R}^{(n{-}r) \times n}$ whose rows are a basis for the left syzygy of $M^{\star}N{+}N^{\star}M$.
\item Condition \ref{nl:prpc3c3} of Lemma \ref{lem:prpc3ec} holds if and only if $V(M^{\star}N{+}N^{\star}M)$ is divisible on the right by $F$.
\end{enumerate}
\item Using the matrices $P$ and $Q$:
\begin{enumerate}
\item Compute a $V \in \mathbb{R}^{(n{-}r) \times n}$ whose rows are a basis for the left syzygy of $PQ^{\star} {+} QP^{\star}$.
\item Condition \ref{nl:prpc3c5} of Lemma \ref{lem:prpc3ec} holds if and only if $V(\lambda)[P \hspace{0.25cm} {-}Q](\lambda)$ has full row rank for all $\lambda \in \mathbb{C}$.
\end{enumerate}
\end{enumerate} 
}
\end{rem}

\begin{lem}
\label{lem:prln}
Let $\mathcal{B}$ be as in (\ref{eq:topbd}). If $\mathcal{B}$ is passive, then $(P, Q)$ is a positive-real pair.
\end{lem}

\begin{pf}
That $(P, Q)$ satisfy condition \ref{nl:prpc2} in Definition \ref{def:prp} follows from condition \ref{nl:prpcb2} of Lemma \ref{lem:irprpc}, by noting from (\ref{eq:dca1})--(\ref{eq:dca2}) that $[P \hspace{0.25cm} {-}Q]\text{col}(X \hspace{0.15cm} Y) = F$. Then condition \ref{nl:prpc3} in Definition \ref{def:prp} follows from Lemmas \ref{lem:irprpc}--\ref{lem:prpc3ec}. It remains to show that condition \ref{nl:prpc1} of Definition \ref{def:prp} holds. To see this, first note from  condition \ref{nl:prpcb1} of Lemma \ref{lem:irprpc} that $((M+N)(\bar{\lambda})^{T}(M+N)(\lambda) - (N-M)(\bar{\lambda})^{T}(N-M)(\lambda)) \geq 0$ for all $\lambda \in \overbar{\mathbb{C}}_{+}$. Next, let $\lambda \in \overbar{\mathbb{C}}_{+}$ and $\mathbf{z} \in \mathbb{C}^{n}$ satisfy $(M+N)(\lambda)\mathbf{z} = 0$. Then $-\bar{\mathbf{z}}^{T}((N-M)(\bar{\lambda})^{T}(N-M)(\lambda))\mathbf{z} \geq 0$, which implies that $(N-M)(\lambda)\mathbf{z} = 0$, and so $M(\lambda)\mathbf{z} = 0$ and $N(\lambda)\mathbf{z} = 0$. Then from (\ref{eq:dca2}) we obtain $\mathbf{z} = U(\lambda)M(\lambda)\mathbf{z} + V(\lambda)N(\lambda)\mathbf{z} = 0$. We conclude that $(M+N)(\lambda)$ is nonsingular for all $\lambda \in \overbar{\mathbb{C}}_{+}$. Accordingly, with the notation $H \coloneqq (N-M)(M+N)^{-1}$, then $I \geq H(\bar{\lambda})^{T}H(\lambda)$ for all $\lambda \in \overbar{\mathbb{C}}_{+}$. This implies that $I \geq H(\lambda)H(\bar{\lambda})^{T}$ for all $\lambda \in \overbar{\mathbb{C}}_{+}$ (to see this, let $X = H(\lambda)$, and note that $I-X\bar{X}^{T} = (I-X\bar{X}^{T})(I-X\bar{X}^{T})+X(I-\bar{X}^{T}X)\bar{X}^{T}$). Then, noting that $PM = QN$ implies that $(P+Q)H = (P-Q)(M+N)(M+N)^{-1} = P-Q$, we find that $(P+Q)(\lambda)(P+Q)(\bar{\lambda})^{T} \geq (P+Q)(\lambda)H(\lambda)H(\bar{\lambda})(P+Q)(\bar{\lambda})^{T} = (P-Q)(\lambda)(P-Q)(\bar{\lambda})^{T}$ for all $\lambda \in \overbar{\mathbb{C}}_{+}$. We conclude that condition \ref{nl:prpc1} of Definition \ref{def:prp} holds. \qed
\end{pf}

\section{Passive behavior theorem}
\label{sec:pbtp}

In this final section, we prove Theorems \ref{thm:pbtp2} and \ref{thm:pbt}.

{\bf PROOF OF THEOREM \ref{thm:pbt}} (see p.\ \pageref{thm:pbt}). We first prove that \ref{nl:pbtc1} $\Rightarrow$ \ref{nl:pbtc2} $\Rightarrow$ \ref{nl:pbtc4} $\Rightarrow$ \ref{nl:pbtc3} $\Rightarrow$ \ref{nl:pbtc1}. 

\ref{nl:pbtc1} $\Rightarrow$ \ref{nl:pbtc2}. \hspace{0.3cm} By Lemma \ref{lem:sstkd}, $\tilde{\mathcal{B}}$ takes the form of (\ref{eq:topbdpr}). Hence, $(\tilde{P}, \tilde{Q})$ is a positive-real pair by Lemma \ref{lem:prln}. 

\ref{nl:pbtc2} $\Rightarrow$ \ref{nl:pbtc4}. \hspace{0.3cm} To prove this implication, we will show conditions \ref{nl:ausc1} and \ref{nl:ausc2} below. The notation in those conditions is as follows. We let $T = \text{col}(T_{1} \hspace{0.15cm} T_{2})$ be such that $\tilde{C} = [\tilde{C}_{1} \hspace{0.15cm} 0] = CT^{-1}$ and $\tilde{A} = TAT^{-1}$ have the observer staircase form indicated in note \ref{nl:pis5}, and we let $TB \eqqcolon \tilde{B}$ and $T_{1}B \eqqcolon \tilde{B}_{1}$. Then, with $\tilde{A}_{11} \in \mathbb{R}^{d_{1} \times d_{1}}$ as in note \ref{nl:pis5}, we let $\tilde{T} \in \mathbb{R}^{d_{1} \times d_{1}}$ be such that $\tilde{T}\tilde{A}_{11}\tilde{T}^{-1} = \text{diag}(A_{s} \hspace{0.15cm} A_{u})$, where $\text{spec}(A_{s}) \in \mathbb{C}_{-}$ and $\text{spec}(A_{u}) \in \overbar{\mathbb{C}}_{+}$ \citep[Chapter VII]{Gant1}.\footnote{This can alternatively be shown using the real Jordan form. Here, letting $d_{s}$ denote the number of columns (and rows) of $A_{s}$, then the first $d_{s}$ rows (resp., last $d_{1} - d_{s}$ rows) of $\tilde{T}$ span the stable (resp., unstable) left eigenspace of $\tilde{A}_{11}$.} We partition $\tilde{T}\tilde{B}_{1}$ and $\tilde{C}_{1}\tilde{T}^{-1}$ compatibly with $\tilde{T}\tilde{A}_{11}\tilde{T}^{-1} = \text{diag}(A_{s} \hspace{0.15cm} A_{u})$ as $\tilde{T}\tilde{B}_{1} = \text{col}(B_{s} \hspace{0.15cm} B_{u})$ and $\tilde{C}_{1}\tilde{T}^{-1} = [C_{s} \hspace{0.15cm} C_{u}]$. We then let $G_{s}(\xi) = D + C_{s}(\xi I {-} A_{s})^{-1}B_{s}$ and $G_{u}(\xi) = C_{u}(\xi I {-} A_{u})^{-1}B_{u}$, and direct calculation shows that $G(\xi) = D + \tilde{C}_{1}(\xi I - \tilde{A}_{11})^{-1}\tilde{B}_{1} = G_{s}(\xi) + G_{u}(\xi)$. We will show the following.
\begin{enumerate}[label=(\roman*)]
\item There exists a real $X_{u} > 0$ such that $-A_{u}^{T}X_{u} - X_{u}A_{u} = 0$ and $C_{u}^{T} - X_{u}B_{u} = 0$. \label{nl:ausc1}
\item There exist real matrices $X_{s}, L, W$ such that $X_{s} > 0$, $-A_{s}^{T}X_{s} - X_{s}A_{s} = L^{T}L$, $C_{s}^{T} - X_{s}B_{s} = L^{T}W$, and $D + D^{T} = W^{T}W$, where $W + L(\xi I {-} A_{s})^{-1}B_{s}$ is a spectral factor of $G + G^{\star}$. \label{nl:ausc2}
\end{enumerate}

We note that $\hat{T} \coloneqq \text{col}(\tilde{T}T_{1} \hspace{0.15cm} T_{2}) = \text{diag}(\tilde{T} \hspace{0.15cm} I)T$ is nonsingular. Then, with $\hat{A} \coloneqq \hat{T}A\hat{T}^{-1}$, $\hat{B} \coloneqq \hat{T}B$, $\hat{C} \coloneqq C\hat{T}^{-1}$, $\hat{X} \coloneqq \text{diag}(X_{s} \hspace{0.15cm} X_{u} \hspace{0.15cm} 0)$, $L_{\hat{X}} \coloneqq [L \hspace{0.15cm} 0 \hspace{0.15cm} 0]$, and $W_{\hat{X}} \coloneqq W$, it can be verified that $\hat{X} \geq 0$; $-\hat{A}^{T}\hat{X} - \hat{X}\hat{A} = \text{diag}(({-}A_{s}^{T}X_{s} {-} X_{s}A_{s}) \hspace{0.15cm} ({-}A_{u}^{T}X_{u} {-} X_{u}A_{u}) \hspace{0.15cm} 0) = L_{\hat{X}}^{T}L_{\hat{X}}$, $\hat{C}^{T} - \hat{X}\hat{B} = \text{col}((C_{s}^{T} {-} X_{s}B_{s}) \hspace{0.15cm} (C_{u}^{T} {-} X_{u}B_{u}) \hspace{0.15cm} 0) =  L_{\hat{X}}^{T}W_{\hat{X}}$ and $D + D^{T} = W_{\hat{X}}^{T}W_{\hat{X}}$; and $Z_{\hat{X}}(\xi) = W_{\hat{X}} {+}  L_{\hat{X}}(\xi I {-} \hat{A})^{-1}\hat{B}$ is a spectral factor of $G {+} G^{\star}$. Finally, with $X {\coloneqq} \hat{T}^{T}\hat{X}\hat{T}, L_{X} {\coloneqq} L_{\hat{X}}\hat{T}$, and $W_{X} {\coloneqq} W_{\hat{X}}$, it can be verified that $X, L_{X}$, and $W_{X}$ satisfy condition \ref{nl:pbtc4}.

We first prove \ref{nl:ausc1}. Direct calculation verifies that $G = \tilde{Q}^{-1}\tilde{P}$. Since $(\tilde{P}, \tilde{Q})$ is a positive-real pair and $\tilde{Q}$ is nonsingular, then $G$ is PR. To see this, note that if $G$ is analytic in $\mathbb{C}_{+}$ then $G(\lambda) + G(\bar{\lambda})^{T} = \tilde{Q}^{-1}(\lambda)(\tilde{P}(\lambda)\tilde{Q}(\bar{\lambda})^{T} + \tilde{Q}(\lambda)\tilde{P}(\bar{\lambda})^{T})(\tilde{Q}^{-1})(\bar{\lambda})^{T} \geq 0$ for all $\lambda \in \mathbb{C}_{+}$, so $G$ is PR. But suppose instead that $G$ has a pole at some $\lambda \in \mathbb{C}_{+}$. By considering the Laurent series for $G$ about $\lambda$, it can be shown that, for any $\epsilon > 0$, there exists $\mathbf{z} \in \mathbb{C}^{n}$ and an $\eta \in \mathbb{C}$ with $|\eta| \leq \epsilon$ such that $\bar{\mathbf{z}}^{T}(G(\lambda + \eta) + G(\bar{\lambda} + \bar{\eta})^{T})\mathbf{z} < 0$: a contradiction. 

Since $G$ is analytic in $\mathbb{C}_{+}$ and $G = G_{u} + G_{s}$ with $G_{s}(\xi) = D + C_{s}(\xi I {-} A_{s})^{-1}B_{s}$ (whose poles are all in $\mathbb{C}_{-}$) and $G_{u}(\xi) = C_{u}(\xi I {-} A_{u})^{-1}B_{u}$ (whose poles are all in $\overbar{\mathbb{C}}_{+}$), then the poles of $G_{u}$ must all be on the imaginary axis. Since, in addition, $G$ is PR, then $G_{u}$ and $G_{s}$ are both PR and $G_{u} + G_{u}^{\star} = 0$ \citep[Section 5.1]{AndVong}. Next, note that $\tilde{\mathcal{B}} \coloneqq \mathcal{B}_{s}^{(\mathbf{u}, \mathbf{y})}$ is stabilizable (by condition \ref{nl:prpc2} of Definition \ref{def:prp}), and has the observable realization in note \ref{nl:pis6}, whence $[\lambda I {-} \tilde{A}_{11} \hspace{0.15cm} \tilde{B}_{1}]$ has full row rank for all $\lambda \in \overbar{\mathbb{C}}_{+}$ (this follows from note \ref{nl:pis8}). It is then easily shown that $[\lambda I {-} A_{u} \hspace{0.15cm} B_{u}]$ has full row rank for all $\lambda \in \mathbb{C}$, so $(A_{u}, B_{u})$ is controllable. Similarly, it can be shown that $(C_{u}, A_{u})$ is observable since $(\tilde{C}_{1}, \tilde{A}_{11})$ is. Thus, $G_{u}(\xi) = C_{u}(\xi I {-} A_{u})^{-1}B_{u}$ is PR with $G_{u} + G_{u}^{\star} = 0$ and with $(A_{u}, B_{u})$ controllable and $(C_{u}, A_{u})$ observable, and so \ref{nl:ausc1} holds by \citep[Theorem 5]{JWDSP2}. 

Next, let $\mathcal{A}_{s}(\xi) := \xi I - A_{s}$; let $M$ and $N$ be as in Lemma \ref{lem:dcab} (so, in particular, $M$ is invertible, and $NM^{-1} = Q^{-1}P = G$, which is PR); let $r = \text{normalrank}(M^{\star}N+N^{\star}M)$; and let $K \in \mathbb{R}^{r \times n}[\xi]$ be a spectral factor for $M^{\star}N+N^{\star}M$ (i.e., $K(\lambda)$ has full row rank for all $\lambda \in \mathbb{C}_{+}$, and $K^{\star}K = M^{\star}N+N^{\star}M$). To prove condition \ref{nl:ausc2}, we will show the following four conditions.

\begin{enumerate}[label=(\alph*), leftmargin=0.55cm]
\item There exist $J \in \mathbb{R}^{n \times d_{s}}[\xi]$ and $L \in \mathbb{R}^{r \times d_{s}}$ such that $K^{\star}L + J\mathcal{A}_{s} = M^{\star}C_{s}$.\label{nl:pbtci}
\item With $L$ as in \ref{nl:pbtci}, there exists $X_{s} \in \mathbb{R}^{d_{s} \times d_{s}}$ such that $-A_{s}^{T}X_{s} - X_{s}A_{s} = L^{T}L$.\label{nl:pbtcii}
\item $Z \coloneqq KM^{-1}$ is a spectral factor of $G + G^{\star}$ and, with $W \coloneqq \lim_{\xi \rightarrow \infty}Z(\xi)$, then $Z = W + L\mathcal{A}_{s}^{-1}B_{s}$. In particular, $D + D^{T} = W^{T}W$.\label{nl:pbtciii}
\item With $Z, W, L$ as in \ref{nl:pbtci}--\ref{nl:pbtciii}, then $C_{s}^{T}{-}X_{s}B_{s} {=} L^{T}W$.\label{nl:pbtciv}
\end{enumerate}
To show \ref{nl:pbtci}, recall that $K^{\star} \in \mathbb{R}^{n \times r}[\xi]$ satisfies $\text{normalrank}(K^{\star}) = r$, 
and let $\text{col}(H_{1} \hspace{0.15cm} H_{2}) = H \in \mathbb{R}^{n \times n}[\xi]$ be a unimodular matrix such that the rows of $H_{2} \in \mathbb{R}^{(n{-}r) \times n}[\xi]$ are a basis for the left syzygy of $K^{\star}$ (e.g., consider the upper echelon form for $K^{\star}$, see note \ref{nl:pm3}). Then $H_{2}K^{\star} = 0$, $H_{1}K^{\star} \in \mathbb{R}^{r \times r}[\xi]$, and $\text{normalrank}(H_{1}K^{\star}) = r$. We will show that: (a)(i) there exists $L \in \mathbb{R}^{r \times d_{s}}$ and $J_{1} \in \mathbb{R}^{r \times d_{s}}[\xi]$ such that $H_{1}K^{\star}L + J_{1}\mathcal{A}_{s} = H_{1}M^{\star}C_{s}$; and (a)(ii) there exists $J_{2} \in \mathbb{R}^{(n{-}r) \times d_{s}}[\xi]$ such that $J_{2}\mathcal{A}_{s} = H_{2}M^{\star}C_{s}$. Since $H_{2}K^{\star} = 0$ and $H$ is unimodular, then $J {:=} H^{-1}\text{col}(J_{1} \hspace{0.15cm} J_{2})$ and $L$ as above satisfy condition \ref{nl:pbtci}.

To see (a)(i), note from the definitions of $H$, $K$ and $\mathcal{A}_{s}$ that $(H_{1}K^{\star})(\lambda)$ is nonsingular for all $\lambda \in \mathbb{C}_{-}$ and $\mathcal{A}_{s}(\lambda)$ is nonsingular for all $\lambda \in \overbar{\mathbb{C}}_{+}$. Furthermore, from \citep[pp.\ 77--79]{Gant1}, there exist $E \in \mathbb{R}^{r \times d_{s}}[\xi]$ and $F \in \mathbb{R}^{r \times d_{s}}$ such that $H_{1}M^{\star}C_{s} = E\mathcal{A}_{s} + F$. Then, from \citep[Theorem II]{MPEFBN}, there exist $L \in \mathbb{R}^{r \times d_{s}}$ and $R \in \mathbb{R}^{r \times d_{s}}[\xi]$ such that $H_{1}K^{\star}L + R\mathcal{A}_{s} = F$. With $J_{1} := E+R$, we obtain condition (a)(i).

To see (a)(ii), let $\hat{\mathcal{B}}_{s} := \lbrace (\mathbf{y}, \mathbf{x}_{s}) \in \mathcal{L}_{1}^{\text{loc}}\left(\mathbb{R}, \mathbb{R}^{n}\right) \times \mathcal{L}_{1}^{\text{loc}}(\mathbb{R}, \mathbb{R}^{d_{s}}) \mid \tfrac{d\mathbf{x}_{s}}{dt} = A_{s}\mathbf{x}_{s} \text{ and } \mathbf{y} = C_{s}\mathbf{x}_{s}\rbrace$. Note that, if $(\mathbf{v}_{s}, \mathbf{x}_{s}) \in \hat{\mathcal{B}}_{s}$, then $\mathbf{v}_{s} \in \mathcal{E}_{\mathbb{C}_{-}}\left(\mathbb{R}, \mathbb{R}^{n}\right)$ and $(0, \mathbf{v}_{s}, \hat{T}^{-1}\text{col}(\mathbf{x}_{s} \hspace{0.15cm} 0 \hspace{0.15cm} 0)) \in \mathcal{B}_{s}$, and so $(0, \mathbf{v}_{s}) \in \tilde{\mathcal{B}}  \cap \mathcal{E}_{\mathbb{C}_{-}}\left(\mathbb{R}, \mathbb{R}^{n}\right) \times \mathcal{E}_{\mathbb{C}_{-}}\left(\mathbb{R}, \mathbb{R}^{n}\right)$. Then, let $U \in \mathbb{R}^{n \times n}[\xi]$ and $V \in \mathbb{R}^{n \times d_{s}}[\xi]$ be left coprime matrices satisfying $UC_{s} = V\mathcal{A}_{s}$, so, from \citet{JW_TS, PRSMLS, THBRSF}, it follows that $\hat{\mathcal{B}}_{s}^{(\mathbf{y})} = \lbrace \mathbf{y} \in \mathcal{L}_{1}^{\text{loc}}\left(\mathbb{R}, \mathbb{R}^{n}\right) \mid U(\tfrac{d}{dt})\mathbf{y} = 0\rbrace$ (c.f., Lemma \ref{lem:sstkd}). Thus, if $\mathbf{v}_{s} \in \mathcal{L}_{1}^{\text{loc}}\left(\mathbb{R}, \mathbb{R}^{n}\right)$ satisfies $U(\tfrac{d}{dt})\mathbf{v}_{s} = 0$, then $(0, \mathbf{v}_{s}) \in \tilde{\mathcal{B}}  \cap \mathcal{E}_{\mathbb{C}_{-}}\left(\mathbb{R}, \mathbb{R}^{n}\right) \times \mathcal{E}_{\mathbb{C}_{-}}\left(\mathbb{R}, \mathbb{R}^{n}\right)$. Next, note from Lemmas \ref{lem:prpc3ec} and \ref{lem:prln} that condition \ref{nl:prpcb3} of Lemma \ref{lem:irprpc} holds. Also, since $H_{2}K^{\star} = 0$, then $H_{2}K^{\star}K = H_{2}(M^{\star}N+N^{\star}M) = 0$. Thus, if $\mathbf{v}_{s} \in \mathcal{L}_{1}^{\text{loc}}\left(\mathbb{R}, \mathbb{R}^{n}\right)$ satisfies $U(\tfrac{d}{dt})\mathbf{v}_{s} = 0$, then $H_{2}(\tfrac{d}{dt})M^{\star}(\tfrac{d}{dt})\mathbf{v}_{s} = 0$. It follows from note \ref{nl:bt2b} that there exists $S \in \mathbb{R}^{(n{-}r) \times n}[\xi]$ such that $H_{2}M^{\star} = SU$, whence $H_{2}M^{\star}C_{s} = SUC_{s} = SV\mathcal{A}_{s}$. With $J_{2} := SV$, we obtain condition (a)(ii), which completes the proof of condition (a).

Condition \ref{nl:pbtcii} follows as $\text{spec}(A_{s}) {\in} \mathbb{C}_{-}$ implies that $X_{s} = \smallint_{0}^{\infty}e^{A_{s}^{T}t}L^{T}Le^{A_{s}t}dt \geq 0$ satisfies ${-}A_{s}^{T}X_{s} {-} X_{s}A_{s} {=} L^{T}L$. 

To show \ref{nl:pbtciii}, we first let $\lambda \in \mathbb{C}_{+}$ and $\mathbf{z} \in \mathbb{C}^{n}$ satisfy $M(\lambda)\mathbf{z} = 0$. We recall that $NM^{-1} = G$ is PR, so $G$ is analytic in $\mathbb{C}_{+}$, whence $N(\lambda)\mathbf{z} = G(\lambda)M(\lambda)\mathbf{z} = 0$. Then, from (\ref{eq:dca2}), it follows that $(U(\lambda)M(\lambda) + V(\lambda)N(\lambda))\mathbf{z} = \mathbf{z} = 0$. We conclude that $M(\lambda)$ is nonsingular for all $\lambda \in \mathbb{C}_{+}$. Since, in addition, $K$ is a spectral factor of $MN^{\star} + NM^{\star}$, then it is straightforward to show that $Z$ is a spectral factor of $G + G^{\star} = M^{-1}N + N^{\star}(M^{-1})^{\star}$.

We next let $W \coloneqq \lim_{\xi \rightarrow \infty}Z(\xi)$, and we will show that: (c)(i) $W + L\mathcal{A}_{s}^{-1}B_{s} - Z$ has no poles in $\mathbb{C}_{-}$; and (c)(ii) $W + L\mathcal{A}_{s}^{-1}B_{s} - Z$ has no poles in $\overbar{\mathbb{C}}_{+}$. Since, in addition, $W = \lim_{\xi \rightarrow \infty}(Z(\xi))$, then $W + L\mathcal{A}_{s}^{-1}B_{s} = Z$. It then follows that $W^{T}W = \lim_{\xi \rightarrow \infty}(Z^{\star}(\xi)Z(\xi)) = \lim_{\xi \rightarrow \infty}(G(\xi) + G^{\star}(\xi)) = D+D^{T}$.

To show (c)(i), we note that $K^{\star}(W + L\mathcal{A}_{s}^{-1}B_{s} - Z) = K^{\star}W + K^{\star}L\mathcal{A}_{s}^{-1}B_{s} - M^{\star}Z^{\star}Z = K^{\star}W + K^{\star}L\mathcal{A}_{s}^{-1}B_{s} - M^{\star}(D + D^{T} + C_{s}\mathcal{A}_{s}^{-1}B_{s} + B_{s}^{T}(\mathcal{A}_{s}^{\star})^{-1}C_{s}^{T})$. Clearly, $(\mathcal{A}_{s}^{\star})^{-1}$ has no poles in $\overbar{\mathbb{C}}_{-}$, and from \ref{nl:pbtci} it follows that  $K^{\star}L\mathcal{A}_{s}^{-1} - M^{\star}C_{s}\mathcal{A}_{s}^{-1} = -J$, which has no poles in $\mathbb{C}_{-}$. Thus, $K^{\star}(W + L\mathcal{A}_{s}^{-1}B_{s} - Z)$ has no poles in $\mathbb{C}_{-}$. Since $K^{\star}(\lambda)$ has full column rank for all $\lambda \in \mathbb{C}_{-}$, then we conclude that $W + L\mathcal{A}_{s}^{-1}B_{s} - Z$ has no poles in $\mathbb{C}_{-}$. 

To see (c)(ii), we note that, since $G + G^{\star} = D + D^{T} + C_{s}\mathcal{A}_{s}^{-1}B_{s} + B_{s}^{T}(\mathcal{A}_{s}^{\star})^{-1}C_{s}^{T}$, and $\mathcal{A}_{s}^{-1}$ (resp., $(\mathcal{A}_{s}^{\star})^{-1}$) has no poles in $\overbar{\mathbb{C}}_{+}$ (resp., $\overbar{\mathbb{C}}_{-}$), then $(G + G^{\star})(j\omega)$ is analytic for all $\omega \in \mathbb{R}$, whence $Z$ is analytic in $\overbar{\mathbb{C}}_{+}$. It follows that $W + L\mathcal{A}_{s}^{-1}B_{s} - Z$ has no poles in $\overbar{\mathbb{C}}_{+}$. This completes the proof of condition (c).

Finally, to show condition \ref{nl:pbtciv}, note initially from \ref{nl:pbtcii} that $X_{s}\mathcal{A}_{s}^{-1} + (\mathcal{A}_{s}^{\star})^{-1}X_{s} = (\mathcal{A}_{s}^{\star})^{-1}L^{T}L\mathcal{A}_{s}^{-1}$. Next, note that $M^{\star}(W^{T}L + B_{s}^{T}X_{s} - C_{s})\mathcal{A}_{s}^{-1} = M^{\star}(W^{T} + B_{s}^{T}(\mathcal{A}_{s}^{\star})^{-1}L^{T})L\mathcal{A}_{s}^{-1} - M^{\star}C_{s}\mathcal{A}_{s}^{-1} - M^{\star}B_{s}^{T}(\mathcal{A}_{s}^{\star})^{-1}X_{s}$. Also, $M^{\star}(W^{T} + B_{s}^{T}(\mathcal{A}_{s}^{\star})^{-1}L^{T}) = M^{\star}Z^{\star} = K^{\star}$ by \ref{nl:pbtciii}. Thus, from \ref{nl:pbtci}, we find that  $M^{\star}(W^{T}L + B_{s}^{T}X_{s} - C_{s})\mathcal{A}_{s}^{-1} = (K^{\star}L - M^{\star}C_{s})\mathcal{A}_{s}^{-1} - M^{\star}B_{s}^{T}(\mathcal{A}_{s}^{\star})^{-1}X_{s} = -J - M^{\star}B_{s}^{T}(\mathcal{A}_{s}^{\star})^{-1}X_{s}$. It follows that $M^{\star}(W^{T}L + B_{s}^{T}X_{s} - C_{s})\mathcal{A}_{s}^{-1}$ has no poles in $\mathbb{C}_{-}$. But $M^{\star}(\lambda)$ is nonsingular for all $\lambda \in \mathbb{C}_{-}$, and we conclude that $(W^{T}L + B_{s}^{T}X_{s} - C_{s})\mathcal{A}_{s}^{-1}$ has no poles in $\mathbb{C}_{-}$. It is then straightforward to show that $W^{T}L + B_{s}^{T}X_{s} - C_{s} = 0$.

\ref{nl:pbtc4} $\Rightarrow$ \ref{nl:pbtc3}.  \hspace{0.3cm} Immediate.

\ref{nl:pbtc3} $\Rightarrow$ \ref{nl:pbtc1}. \hspace{0.3cm} Consider a fixed but arbitrary $(\mathbf{u}, \mathbf{y}, \mathbf{x})  \in \mathcal{B}_{s}$ and $t_{0} \in \mathbb{R}$, and let $(\mathbf{\hat{u}}, \mathbf{\hat{y}}) \in \tilde{\mathcal{B}}= \mathcal{B}_{s}^{(\mathbf{u}, \mathbf{y})}$ satisfy $\mathbf{\hat{u}}(t) = \mathbf{u}(t)$ and $\mathbf{\hat{y}}(t) = \mathbf{y}(t)$ for all $t < t_{0}$. Then, from note \ref{nl:pis6}, there exists $(\mathbf{\hat{u}}, \mathbf{\hat{y}}, \mathbf{\hat{x}}) \in \mathcal{B}_{s}$ with $\mathbf{\hat{x}}(t_{0}) = \mathbf{x}(t_{0})$. From Remark \ref{rem:sfd}, since $\hat{\mathbf{x}}^{T}(t_{1})X\hat{\mathbf{x}}(t_{1}) \geq 0$ and $\mathbf{\hat{x}}(t_{0}) = \mathbf{x}(t_{0})$, then $-\smallint_{t_{0}}^{t_{1}}\mathbf{\hat{u}}^{T}(t)\mathbf{\hat{y}}(t) dt \leq \tfrac{1}{2} \mathbf{x}^{T}(t_{0})X\mathbf{x}(t_{0})$. This inequality holds for all $(\mathbf{\hat{u}}, \mathbf{\hat{y}}) \in \tilde{\mathcal{B}}$ that satisfy $(\mathbf{\hat{u}}(t), \mathbf{\hat{y}}(t)) = (\mathbf{u}(t), \mathbf{y}(t))$ for all $t < t_{0}$, so $\tilde{\mathcal{B}}$ is passive.

We next assume that $D+D^{T} > 0$, and we prove that \ref{nl:pbtc4} $\Rightarrow$ \ref{nl:pbtc5} $\Rightarrow$ \ref{nl:pbtc3}. First, let $X, L_{X}, W_{X}$ and $Z_{X}$ be as in condition \ref{nl:pbtc4}. Since $n \geq \text{normalrank}(G+G^{\star}) \geq \text{rank}(D+D^{T}) = n$, then $\text{normalrank}(G+G^{\star}) = n$, so $Z_{X} \in \mathbb{R}^{n \times n}(\xi)$, and $W_{X} = \lim_{\xi \rightarrow \infty}(Z(\xi)) \in \mathbb{R}^{n \times n}$. As $W_{X}^{T}W_{X} = D+D^{T}$, which is nonsingular, then $W_{X}$ is nonsingular. We then find that $-A^{T}X-XA - (C^{T}-XB)(D+D^{T})^{-1}(C-B^{T}X) = L_{X}^{T}L_{X} - L_{X}^{T}W_{X}W_{X}^{-1}(W_{X}^{T})^{-1}W_{X}^{T}L_{X} = 0$. Next, suppose $X \geq 0$ is real and satisfies $\Pi(X) = 0$; let $W_{X}$ be a real nonsingular matrix with $D+D^{T} = W_{X}^{T}W_{X}$; and let $L_{X} \coloneqq (W_{X}^{T})^{-1}(C-B^{T}X)$. Then $X, L_{X}$ and $W_{X}$ satisfy condition \ref{nl:pbtc3}.

We now prove condition \ref{nl:pbt2ac1}. Accordingly, suppose condition \ref{nl:pbtc3} holds, and let $X, L_{X}$ and $W_{X}$ be as in that condition. To show condition \ref{nl:pbt2ac1}(a), suppose that $(C,A)$ is observable and there exists $\mathbf{z} \in \mathbb{R}^{d}$ with $X\mathbf{z} = 0$. Since $X$ is symmetric, then $\mathbf{z}^{T}X = 0$. Thus, $\mathbf{z}^{T}(-A^{T}X-XA)\mathbf{z} = \mathbf{z}^{T}L_{X}^{T}L_{X}\mathbf{z} = 0$, whence $L_{X}\mathbf{z} = 0$. It follows that $(C-B^{T}X)\mathbf{z} = W_{X}^{T}L_{X}\mathbf{z} = 0$, so $C\mathbf{z} = 0$. Also,  $(-A^{T}X-XA)\mathbf{z} = L_{X}^{T}L_{X}\mathbf{z} = 0$, so $XA\mathbf{z} = 0$. By replacing $\mathbf{z}$ with $A\mathbf{z}$ in the preceding argument, we find that $L_{X}A\mathbf{z} = 0$, $CA\mathbf{z} = 0$, and $XA^{2}\mathbf{z} = 0$. Proceeding inductively gives $CA^{k}\mathbf{z} = 0$ ($k = 0, 1, 2, \ldots$). Since $(C,A)$ is observable, then $\mathbf{z} = 0$, and we conclude that $X > 0$. To show condition \ref{nl:pbt2ac1}(b), let $\lambda \in \mathbb{C}_{+}$ and $\mathbf{z} \in \mathbb{C}^{d}$ satisfy $(\lambda I {-} A)\mathbf{z} = 0$. Then $\mathbf{\bar{z}}^{T}L_{X}^{T}L_{X}\mathbf{z} = \mathbf{\bar{z}}^{T}(-A^{T}X-XA)\mathbf{z} = -(\bar{\lambda} + \lambda)\bar{\mathbf{z}}^{T}X\mathbf{z} \leq 0$, whence $L_{X}\mathbf{z} = 0$ and $\bar{\mathbf{z}}^{T}X\mathbf{z} = 0$. Since $X > 0$ by condition \ref{nl:pbt2ac1}(a), then $\mathbf{z} = 0$, and we conclude that $\text{spec}(A) \in \overbar{\mathbb{C}}_{-}$.

It remains to prove condition \ref{nl:pbt2ac2}. Condition \ref{nl:pbt2ac2}(a) was shown in the proof of \ref{nl:pbtc4} $\Rightarrow$ \ref{nl:pbtc5}. To see condition \ref{nl:pbt2ac2}(b), note that $A+B(D+D^{T})^{-1}(B^{T}X - C) = A - BW_{X}^{-1}L_{X}$, and consider a fixed but arbitrary $\lambda \in \mathbb{C}_{+}$. From the proof of condition (i)(b), if $\mathbf{z} \in \mathbb{C}^{d}$ satisfies $(\lambda I {-} A)\mathbf{z} = 0$, then $L_{X}\mathbf{z} = 0$, whence $(\lambda I {-} (A - BW_{X}^{-1}L_{X}))\mathbf{z} = (\lambda I {-} A)\mathbf{z} = 0$. It remains to show that if $\lambda I {-} A$ is nonsingular, then $\lambda I {-} (A - BW_{X}^{-1}L_{X})$ is nonsingular. Accordingly, suppose that $\lambda I {-} A$ is nonsingular and $\mathbf{y} \in \mathbb{C}^{d}$ satisfies $\mathbf{y}^{T}(\lambda I {-} (A - BW_{X}^{-1}L_{X})) = 0$. Then $\mathbf{y}^{T}(\lambda I {-} (A - BW_{X}^{-1}L_{X}))(\lambda I {-} A)^{-1}B = \mathbf{y}^{T}B + \mathbf{y}^{T}BW_{X}^{-1}L_{X}(\lambda I {-} A)^{-1}B = \mathbf{y}^{T}B + \mathbf{y}^{T}BW_{X}^{-1}(Z_{X}(\lambda) - W_{X}) = \mathbf{y}^{T}BW_{X}^{-1}Z_{X}(\lambda) = 0$. But $W_{X}^{-1}Z_{X}(\lambda)$ has full row rank, so $\mathbf{y}^{T}B = 0$. Thus, $\mathbf{y}^{T}(\lambda I {-} A) = \mathbf{y}^{T}(\lambda I {-} (A - BW_{X}^{-1}L_{X})) - \mathbf{y}^{T}BW_{X}^{-1}L_{X} = 0$, which implies that $\mathbf{y} = 0$. 
\qed

\begin{rem}
\label{rem:pcd}
\textnormal{
We note that the matrix $L$ in the above theorem can be obtained by considering the Jordan chains of $A_{s}$. Specifically, let $A_{s}$ have eigenvalues $\lambda_{1}, \ldots, \lambda_{n}$ with Jordan chains $(\mathbf{v}_{1,1}, \ldots, \mathbf{v}_{1,N(\lambda_{1})}), \ldots,$ $(\mathbf{v}_{n,1}, \ldots , \mathbf{v}_{n,N(\lambda_{n})})$. Also, for any given $H \in \mathbb{R}^{m \times n}(\xi)$ and $\lambda \in \mathbb{C}$ such that $\lambda$ is not a pole of $H$, let $H_{\lambda, j}$ denote the $j{+}1$th term in the Taylor expansion for $H$ about $\lambda$, i.e., $H_{\lambda, j} = \tfrac{1}{j!}(\tfrac{d^{j}}{d\xi^{j}}H)(\lambda)$ for $j = 0,1, \ldots$. Then $L$ can be obtained by solving the equations
\begin{equation}
\sum_{j=0}^{k-1}(K^{\star}_{\lambda_{i},j}L\mathbf{v}_{i,k-j} - M^{\star}_{\lambda_{i}, j}C_{s}\mathbf{v}_{i,k-j}) = 0,\label{eq:sL}
\end{equation}
for $i = 1, \ldots, n$ and $k = 1, \ldots, N(\lambda_{i})$. It can then be shown that, if $\text{spec}(A) \in \mathbb{C}_{-}$ and condition \ref{nl:pbtc3} of Theorem \ref{thm:pbt} holds, then \citep[equation (4)]{POPRL} must hold.}

\textnormal{To show (\ref{eq:sL}), we let $\mathcal{A}_{s}(\xi) \coloneqq \xi I - A_{s}$, and we consider the Jordan chain for $A_{s}$ corresponding to an eigenvalue $\lambda$: $\mathcal{A}_{s}(\lambda)\mathbf{v}_{1} = 0$, $\mathcal{A}_{s}(\lambda)\mathbf{v}_{j} + \mathbf{v}_{j-1} = 0$ ($j = 2, \ldots , N(\lambda)$). If $J \in \mathbb{R}^{n \times d_{s}}[\xi]$ and $L \in \mathbb{R}^{r \times d_{s}}$ satisfy $K^{\star}L + J\mathcal{A}_{s} = M^{\star}C_{s}$, then $K^{\star}_{\lambda, j}L - M^{\star}_{\lambda, j}C_{s} = {-}\tfrac{1}{j!}\tfrac{d^{j}}{d\xi^{j}}(J\mathcal{A}_{s})(\lambda) = {-}J_{\lambda,j}\mathcal{A}_{s}(\lambda) {-} J_{\lambda,j-1}$ (where $J_{\lambda,-1} \coloneqq 0$). Thus, for $k = 1, \ldots, N(\lambda)$, $\sum_{j=0}^{k-1}(K^{\star}_{\lambda,j}L\mathbf{v}_{k-j} - M^{\star}_{\lambda, j}C_{s}\mathbf{v}_{k-j})$ $= {-}\sum_{j=0}^{k-1}{(J_{\lambda,j}\mathcal{A}_{s}(\lambda)\mathbf{v}_{k-j})}$ ${-} \sum_{i = 1}^{k-1}{(J_{\lambda,i-1}\mathbf{v}_{k-i})}$ $= {-}\sum_{j=0}^{k-2}{(J_{\lambda,j}(\mathcal{A}_{s}(\lambda)\mathbf{v}_{k-j} {+} \mathbf{v}_{k - j-1}))} {-} J_{\lambda,k-1}\mathcal{A}_{s}(\lambda)\mathbf{v}_{1} {=} 0$.
}
\end{rem}

{\bf PROOF OF THEOREM \ref{thm:pbtp2}} (see p.\ \pageref{thm:pbtp2}). That \ref{nl:pbt2c1} $\Rightarrow$ \ref{nl:pbt2c2} was shown in Lemma \ref{lem:prln}. Here, prove that \ref{nl:pbt2c2} $\Rightarrow$ \ref{nl:pbt2c3} $\Rightarrow$ \ref{nl:pbt2c1}. 

\ref{nl:pbt2c2} $\Rightarrow$ \ref{nl:pbt2c3}.  \hspace{0.3cm} Let $\hat{P} \coloneqq P-Q$ and $\hat{Q} \coloneqq P+Q$. Since $(P,Q)$ is a positive-real pair, then $\hat{Q}(\lambda)\hat{Q}(\bar{\lambda})^{T} - \hat{P}(\lambda)\hat{P}(\bar{\lambda})^{T} \geq 0$ and  $\text{rank}([\hat{P} \hspace{0.25cm} {-}\hat{Q}](\lambda)) = n$ for all $\lambda \in \overbar{\mathbb{C}}_{+}$. We will show that: (i) $\hat{Q}(\lambda)$ is nonsingular for all $\lambda \in \overbar{\mathbb{C}}_{+}$; and (ii) $\hat{Q}^{-1}\hat{P}$ is proper. To see (i), suppose $\mathbf{z} \in \mathbb{C}^{n}$ and $\lambda \in \overbar{\mathbb{C}}_{+}$ satisfy $\mathbf{z}^{T}\hat{Q}(\lambda) = 0$. Then $-\mathbf{z}^{T}\hat{P}(\lambda)\hat{P}(\bar{\lambda})^{T}\bar{\mathbf{z}} \geq 0$, which implies that $\mathbf{z}^{T}\hat{P}(\lambda) = 0$. Since  $\text{rank}([\hat{P} \hspace{0.25cm} {-}\hat{Q}](\lambda)) = n$, then this implies that $\mathbf{z} = 0$. To see (ii), note that, since $\hat{Q}(\lambda)$ is nonsingular for all $\lambda \in \overbar{\mathbb{C}}_{+}$, then $I - (\hat{Q}^{-1}\hat{P})(\lambda)(\hat{Q}^{-1}\hat{P})(\bar{\lambda})^{T} \geq 0$ for all $\lambda \in \overbar{\mathbb{C}}_{+}$, and it is then easily shown that $\hat{Q}^{-1}\hat{P}$ is proper.

Let $R \in \mathbb{R}^{m \times n}[\xi]$ with $\text{normalrank}(R) = m$, and recall the notation $\Delta(R)$ from the proof of Lemma \ref{lem:dcab}. If $R$ is partitioned as $R = [R_{1} \hspace{0.25cm} R_{2}]$ where $R_{2} \in \mathbb{R}^{m \times m}[\xi]$ is nonsingular, then $R_{2}^{-1}R_{1}$ is proper if and only if $\deg(\det(R_{2})) = \Delta(R)$ \citep[Theorem 3.3.22]{JWIMTSC}. Thus, $\deg(\det(\hat{Q})) = \Delta([\hat{P} \hspace{0.25cm} {-}\hat{Q}])$. But $\det(\hat{Q}) = \det(P+Q)$, which is the sum of all the determininants composed of columns of $P$ together with the complementary columns of $Q$ (i.e., $\det([\mathbf{p}_{1} \hspace{0.15cm} \mathbf{p}_{2} \hspace{0.15cm} \cdots]) + \det([\mathbf{q}_{1} \hspace{0.15cm} \mathbf{p}_{2} \hspace{0.15cm} \cdots]) + \det([\mathbf{p}_{1} \hspace{0.15cm} \mathbf{q}_{2} \hspace{0.15cm} \cdots]) + \ldots$, where $\mathbf{p}_{k}$ (resp., $\mathbf{q}_{k}$) denotes the $k$th column of $P$ (resp., $Q$)). From among the determinants in this sum, we pick one of greatest degree, we let  $T = \text{col}(T_{1} \hspace{0.25cm} T_{2})$ be a permutation matrix such that $T_{1}^{T}$ (resp., $T_{2}^{T}$) selects the columns from $Q$ (resp., $P$) appearing in this determinant, and we define $\tilde{Q} \coloneqq [QT_{1}^{T} \hspace{0.25cm} {-}PT_{2}^{T}]$ and $\tilde{P} \coloneqq [PT_{1}^{T} \hspace{0.25cm} {-}QT_{2}^{T}]$. Then $\deg(\det(\tilde{Q})) \geq \deg(\det(P+Q)) = \Delta([\hat{P} \hspace{0.25cm} {-}\hat{Q}])$. Furthermore, $T_{1}^{T}T_{1} + T_{2}^{T}T_{2} = I$ as $T$ is a permutation matrix, and with the notation
\begin{equation*}
S_{1} \coloneqq \begin{bmatrix}T_{1}^{T}& 0& 0& T_{2}^{T}\\ 0& T_{2}^{T}& T_{1}^{T}& 0\end{bmatrix}, \text{ and } S_{2} \coloneqq \tfrac{1}{2}\begin{bmatrix}I & I\\ -I& I\end{bmatrix},
\end{equation*}
we find that $S_{1}S_{1}^{T} = 2S_{2}S_{2}^{T} = I$, and $[\tilde{P} \hspace{0.25cm} {-}\tilde{Q}] = [P \hspace{0.25cm} {-}Q]S_{1} = [\hat{P} \hspace{0.25cm} {-}\hat{Q}]S_{2}S_{1}$. Then, from the Binet-Cauchy formula, we obtain $\Delta([\tilde{P} \hspace{0.25cm} {-}\tilde{Q}]) = \Delta([\hat{P} \hspace{0.25cm} {-}\hat{Q}])$ \citep[proof of Theorem 7.2]{THBRSF}. Since, in addition, $\Delta([\tilde{P} \hspace{0.25cm} {-}\tilde{Q}])  \geq \deg(\det(\tilde{Q})) \geq \Delta([\hat{P} \hspace{0.25cm} {-}\hat{Q}])$, then $\deg(\det(\tilde{Q})) = \Delta([\tilde{P} \hspace{0.25cm} {-}\tilde{Q}])$, so $\tilde{Q}^{-1}\tilde{P}$ is proper. 

Since $S_{1}S_{1}^{T} = I$, then $[\tilde{P} \hspace{0.25cm} {-}\tilde{Q}](\tfrac{d}{dt})S_{1}^{T}\text{col}(\mathbf{i} \hspace{0.15cm} \mathbf{v}) = [P \hspace{0.25cm} {-}Q](\tfrac{d}{dt})\text{col}(\mathbf{i} \hspace{0.15cm} \mathbf{v})$. Thus, with $\mathbf{i}_{1} = T_{1}\mathbf{i}$, $\mathbf{i}_{2} = T_{2}\mathbf{i}$, $\mathbf{v}_{1} = T_{1}\mathbf{v}$, and $\mathbf{v}_{2} = T_{2}\mathbf{v}$, it follows that $\mathbf{i}$ and $\mathbf{v}$ have the compatible partitions $\mathbf{i} \coloneqq (\mathbf{i}_{1}, \mathbf{i}_{2})$ and $\mathbf{v} \coloneqq (\mathbf{v}_{1}, \mathbf{v}_{2})$, and $\tilde{\mathcal{B}}\coloneqq \mathcal{B}^{(\text{col}(\mathbf{i}_{1} \hspace{0.15cm} \mathbf{v}_{2}), \text{col}(\mathbf{v}_{1} \hspace{0.15cm} \mathbf{i}_{2}))}$ takes the form of (\ref{eq:topbdpr}). Thus, from Lemma \ref{lem:sstkd}, there exists a state-space system $\mathcal{B}_{s}$ as in (\ref{eq:bne1}) such that $\tilde{\mathcal{B}} = \mathcal{B}_{s}^{(\mathbf{u}, \mathbf{y})}$. Moreover, it is easily verified that $(\tilde{P}, \tilde{Q})$ is a positive-real pair since $(P, Q)$ is, so $\tilde{\mathcal{B}}$ is passive by Theorem \ref{thm:pbt}.

\ref{nl:pbt2c3} $\Rightarrow$ \ref{nl:pbt2c1}.  \hspace{0.3cm} Note from the preceding discussion that $\mathcal{B}$ takes the form of (\ref{eq:topbd}) where $[P \hspace{0.25cm} {-}Q] = [\tilde{P} \hspace{0.25cm} {-}\tilde{Q}]S_{1}^{T}$. Now, let $(\mathbf{i}, \mathbf{v}) \in \mathcal{B}$ and $t_{0} \in \mathbb{R}$; let $(\mathbf{\hat{i}}, \mathbf{\hat{v}}) \in \mathcal{B}$ be a fixed but arbitrary trajectory satisfying  $(\mathbf{\hat{i}}(t), \mathbf{\hat{v}}(t)) = (\mathbf{i}(t), \mathbf{v}(t))$ for all $t < t_{0}$; and let $\mathbf{u} \coloneqq \text{col}(T_{1}\mathbf{i} \hspace{0.2cm} T_{2}\mathbf{v}), \mathbf{y} \coloneqq \text{col}(T_{1}\mathbf{v} \hspace{0.2cm} T_{2}\mathbf{i}), \hat{\mathbf{u}} \coloneqq \text{col}(T_{1}\mathbf{\hat{i}} \hspace{0.2cm} T_{2}\mathbf{\hat{v}})$, and $\hat{\mathbf{y}} \coloneqq \text{col}(T_{1}\mathbf{\hat{v}} \hspace{0.2cm} T_{2}\mathbf{\hat{i}})$.  Then $ (\mathbf{u}, \mathbf{y}) \in \tilde{\mathcal{B}}$, $(\hat{\mathbf{u}}, \hat{\mathbf{y}}) \in \tilde{\mathcal{B}}$, and  $ ( \hat{\mathbf{u}}(t),  \hat{\mathbf{y}}(t)) = (\mathbf{u}(t), \mathbf{y}(t))$ for all $t < t_{0}$. Since $\tilde{\mathcal{B}}$ is passive, there exists a $K \in \mathbb{R}$ such that $-\smallint_{t_{0}}^{t_{1}} \hat{\mathbf{u}}^{T}(t)\hat{\mathbf{y}}(t)dt < K$ for all $t_{1} \geq t_{0}$. Since, in addition $T_{1}^{T}T_{1}+T_{2}^{T}T_{2} = I$, then $-\smallint_{t_{0}}^{t_{1}} \mathbf{\hat{i}}^{T}(t)\mathbf{\hat{v}}(t)dt = -\smallint_{t_{0}}^{t_{1}} \hat{\mathbf{u}}^{T}(t)\hat{\mathbf{y}}(t)dt < K$, and we conclude that $\mathcal{B}$ is passive. \qed

\section{Conclusions}
The positive-real lemma links the concepts of passivity, positive-real transfer functions, spectral factorisation, linear matrix inequalities, and algebraic Riccati equations. However, the lemma only considers systems described by a controllable state-space realization, which leaves important questions unanswered. For example, it does not specify which uncontrollable systems are passive. In this paper, we sought to answer this question and others by proving two new theorems: the passive behavior theorem, parts 1 and 2.

\begin{ack}
This research was conducted in part during a Fellowship supported by the Cambridge Philosophical Society, http://www.cambridgephilosophicalsociety.org.
\end{ack}

\appendix
\section{Polynomial and rational matrices} \label{app:prm}    
Several of the results in this paper depend on the properties of polynomial matrices that we describe here.

\begin{remunerate}
\labitem{\ref{app:prm}\arabic{muni}}{nl:pm1} $U \in \mathbb{R}^{l \times l}[\xi]$ is called \emph{unimodular} if there exists $V \in \mathbb{R}^{l \times l}[\xi]$ such that $UV = I$ (whence $VU = I$). $U$ is unimodular if and only if $\det(U)$ is a non-zero constant.

\labitem{\ref{app:prm}\arabic{muni}}{nl:pm1b} Let $R_{1} \in \mathbb{R}^{l \times n_{1}}[\xi]$ and $R_{2} \in \mathbb{R}^{l \times n_{2}}[\xi]$. We say that $R_{1}$ and $R_{2}$ are \emph{left coprime} if $[R_{1} \hspace{0.25cm} R_{2}](\lambda)$ has full row rank for all $\lambda \in \mathbb{C}$.

\labitem{\ref{app:prm}\arabic{muni}}{nl:pm2} Let $R \in \mathbb{R}^{l \times n}[\xi]$. The \emph{left syzygy} of $R$ is the set of $\mathbf{c} \in \mathbb{R}^{l}[\xi]$ that satisfy $\mathbf{c}^{T}R = 0$. If  $\text{normalrank}(R) = m$, then there exists $V \in \mathbb{R}^{(l{-}m) \times l}[\xi]$ such that (i) $V(\lambda)$ has full row rank for all $\lambda \in \mathbb{C}$; and (ii) $VR = 0$. If $V \in \mathbb{R}^{(l{-}m) \times l}[\xi]$ satisfies (i) and (ii), then $\mathbf{c} \in \mathbb{R}^{l}[\xi]$ is in the left syzygy of $R$ if and only if there exists a $\mathbf{p} \in \mathbb{R}^{l{-}m}[\xi]$ such that $\mathbf{p}^{T}V = \mathbf{c}^{T}$; and we say that the rows of $V$ are a basis for the left syzygy of $R$.

\labitem{\ref{app:prm}\arabic{muni}}{nl:pm3} Given any $R \in \mathbb{R}^{l \times n}[\xi]$ with $\text{normalrank}(R) = m$, there exists a unimodular $U \in \mathbb{R}^{l \times l}[\xi]$ (resp., $V \in \mathbb{R}^{n \times n}[\xi]$) such that $UR = \text{col}(\tilde{R} \hspace{0.15cm} 0_{(l-m) \times n})$ (resp., $RV = [\hat{R} \hspace{0.15cm} 0_{l \times (n - m)}]$), where $\tilde{R} \in \mathbb{R}^{m \times n}[\xi]$ is in either (i) upper echelon form, or (ii) row reduced form (resp., $\hat{R} \in \mathbb{R}^{l \times m}[\xi]$ is in either (ib) lower echelon form, or (iib) column reduced form) (see, e.g., \citet{Gant1} Chapter VI and \citet{WLMS}). The last $l-m$ rows of $U$ are a basis for the left syzygy of $R$. Evidently, if $R$ is para-Hermitian, then $URU^{\star} = \text{diag}(\Phi \hspace{0.15cm} 0)$ where $\Phi \in \mathbb{R}^{m \times m}[\xi]$ is para-Hermitian and nonsingular.
\end{remunerate}

\section{Linear systems and behaviors}
\label{sec:ltidsb}
Here, we provide relevant results from behavioral theory \citep[see][]{JWIMTSC}.
\begin{remunerate}
\labitem{\ref{sec:ltidsb}\arabic{muni}}{nl:bt2} Let $\mathcal{B}_{1} = \lbrace \mathbf{w} \in \mathcal{L}_{1}^{\text{loc}}\left(\mathbb{R}, \mathbb{R}^{k}\right) \mid R_{1}(\tfrac{d}{dt})\mathbf{w} = 0\rbrace$ and $\mathcal{B}_{2} = \lbrace \mathbf{w} \in \mathcal{L}_{1}^{\text{loc}}\left(\mathbb{R}, \mathbb{R}^{k}\right) \mid R_{2}(\tfrac{d}{dt})\mathbf{w} = 0\rbrace$ for some $R_{1}, R_{2} \in \mathbb{R}^{l \times k}[\xi]$. Then $\mathcal{B}_{1} = \mathcal{B}_{2}$ if and only if there exists a unimodular $U \in \mathbb{R}^{l \times l}[\xi]$ such that $R_{1} = UR_{2}$ \citep[Theorem 3.6.2]{JWIMTSC}. The requirement that $R_{1}$ and $R_{2}$ have the same number of rows is of little consequence since the addition or deletion of rows of zeros doesn't alter the behavior.

\labitem{\ref{sec:ltidsb}\arabic{muni}}{nl:bt2b} Let $F \in \mathbb{R}^{m_{1} \times n}[\xi]$, $G \in \mathbb{R}^{m_{2} \times n}[\xi]$, and $\mathcal{B} {\coloneqq} \lbrace \mathbf{z} \in  \mathcal{L}_{1}^{\text{loc}}\left(\mathbb{R}, \mathbb{R}^{n}\right) \mid F(\tfrac{d}{dt})\mathbf{z} = 0\rbrace$. If $\mathbf{z} \in \mathcal{B}$ implies $G(\tfrac{d}{dt})\mathbf{z} = 0$, then there exists $H \in \mathbb{R}^{m_{2} \times m_{1}}[\xi]$ such that $G = HF$. To see this, note that $\mathcal{B} = \lbrace \mathbf{z} \in  \mathcal{L}_{1}^{\text{loc}}\left(\mathbb{R}, \mathbb{R}^{n}\right) \mid \text{col}(F \hspace{0.15cm} G)(\tfrac{d}{dt})\mathbf{z} = 0\rbrace$. It follows from note \ref{nl:bt2} that there exists a unimodular matrix $U$ with $U\text{col}(F \hspace{0.15cm} 0_{m_{2} \times n}) = \text{col}(F \hspace{0.15cm} G)$. We form $H$ from the last $m_{2}$ rows and first $m_{1}$ columns of $U$ to obtain $G = HF$.

\labitem{\ref{sec:ltidsb}\arabic{muni}}{nl:bt3} Consider a system $\mathcal{B}$ as in (\ref{eq:bd}). $\mathcal{B}$ is called \emph{controllable} if, for any two trajectories $\mathbf{w}_{1}, \mathbf{w}_{2} \in \mathcal{B}$ and $t_{0} \in \mathbb{R}$, there exists $\mathbf{w} \in \mathcal{B}$ and $t_{1} \geq t_{0}$ such that $\mathbf{w}(t) = \mathbf{w}_{1}(t)$ for all $t \leq t_{0}$ and $\mathbf{w}(t) = \mathbf{w}_{2}(t)$ for all $t \geq t_{1}$ \citep[Definition 5.2.2]{JWIMTSC}; and \emph{stabilizable} if for any $\mathbf{w}_{1} \in \mathcal{B}$ there exists $\mathbf{w} \in \mathcal{B}$ such that $\mathbf{w}(t) = \mathbf{w}_{1}(t)$ for all $t \leq t_{0}$ and $\lim_{t \rightarrow \infty}\mathbf{w}(t) = 0$ \citep[Definition 5.2.29]{JWIMTSC}. $\mathcal{B}$ is controllable (resp., stabilizable) if and only if the rank of $R(\lambda)$ is the same for all $\lambda \in \mathbb{C}$ (resp., $\lambda \in \overbar{\mathbb{C}}_{+}$) \citep[Theorems 5.2.10, 5.2.30]{JWIMTSC}. 
\end{remunerate}

\section{Bilinear and quadratic differential forms}
\label{sec:bdf}

Bilinear and quadratic differential forms were introduced in \citet{WTQDF}, and are useful for studying dissipativity. Some relevant definitions and results are presented here.

\begin{remunerate}
\labitem{\ref{sec:bdf}\arabic{muni}}{nl:bdf1} A \emph{bilinear differential form} is a mapping from $\mathcal{C}_{\infty}\left(\mathbb{R}, \mathbb{R}^{m}\right) \times \mathcal{C}_{\infty}\left(\mathbb{R}, \mathbb{R}^{n}\right)$ to $\mathcal{C}_{\infty}\left(\mathbb{R}, \mathbb{R}\right)$ of the form $L_{\Phi}(\mathbf{w},\mathbf{x}) \coloneqq \sum_{i=1}^{M}\sum_{j=1}^{N}(\tfrac{d^{i-1}\mathbf{w}}{dt^{i-1}})^{T}\Phi_{ij}(\tfrac{d^{j-1}\mathbf{x}}{dt^{j-1}})$, for some positive integers $M, N$ and $\Phi_{ij} \in \mathbb{R}^{m \times n}$. It is naturally associated with the two variable polynomial matrix $\Phi \in \mathbb{R}^{m \times n}[\xi, \eta]$ defined as $\Phi(\xi, \eta) \coloneqq \sum_{i=1}^{M}\sum_{j=1}^{N}\Phi_{ij}\xi^{i-1}\eta^{j-1}$. If $\Phi \in \mathbb{R}^{m \times m}[\xi, \eta]$, then $Q_{\phi}(\mathbf{w}) {\coloneqq} L_{\phi}(\mathbf{w}, \mathbf{w})$ is called a \emph{quadratic differential form}.

\labitem{\ref{sec:bdf}\arabic{muni}}{nl:bdf2} Let $\Phi \in \mathbb{R}^{m \times n}[\xi, \eta]$ and let $\Psi(\xi, \eta) \coloneqq (\xi + \eta)\Phi(\xi, \eta)$. Then the product rule of differentiation gives $\tfrac{d}{dt}L_{\Phi}(\mathbf{w}, \mathbf{x})  = L_{\Psi}(\mathbf{w},\mathbf{x})$.

\labitem{\ref{sec:bdf}\arabic{muni}}{nl:bdf3} Associated with a given $R \in \mathbb{R}^{m \times n}[\xi]$ is the bilinear differential form $L_{\Phi_{R}}$, where $\Phi_{R}(\xi, \eta) \coloneqq (R(\xi)-R(-\eta))/(\xi + \eta)$. Since $\xi = -\eta$ implies $R(\xi) - R(-\eta) = 0$, then $\Phi_{R} \in \mathbb{R}^{m \times n}[\xi, \eta]$ from the factor theorem. Furthermore, from note \ref{nl:bdf2}, $(R\left(\tfrac{d}{dt}\right)\mathbf{w})^{T}\mathbf{x} - \mathbf{w}^{T} (R^{T}\left(-\tfrac{d}{dt}\right)\mathbf{x}) = \tfrac{d}{dt}L_{\Phi_{R}}(\mathbf{w},\mathbf{x})$, so, for any given $t_{1} \geq t_{0} \in \mathbb{R}$,
\begin{multline*}\hspace*{-0.3cm}\int_{t_{0}}^{t_{1}}{\left(R\left(\tfrac{d}{dt}\right)\mathbf{w}\right)^{T}\!(t)\mathbf{x}(t) dt} \\
\hspace*{-0.3cm}= \int_{t_{0}}^{t_{1}}{\mathbf{w}^{T}(t) \left(\!R^{T}\!\left(\!-\tfrac{d}{dt}\!\right)\!\mathbf{x}\right)\!(t) dt} + \left[L_{\Phi_{R}}(\mathbf{w},\mathbf{x})(t)\right]_{t_{0}}^{t_{1}}.
\end{multline*}
Note that if $R(\tfrac{d}{dt}) = \tfrac{d}{dt}$, then $\Phi_{R} = 1$, and this becomes the formula for integration by parts.
\end{remunerate}

\section{States and state-space systems}
\label{sec:sot}
In this final appendix, we provide several useful definitions and results concerning state-space systems.

\begin{remunerate}
\labitem{\ref{sec:sot}\arabic{muni}}{nl:pis3} Let $\mathcal{B}_{s}$ be as in (\ref{eq:bne1}). Then, for any given $\mathbf{u} \in \mathcal{L}_{1}^{\text{loc}}\left(\mathbb{R}, \mathbb{R}^{n}\right)$, $\mathbf{x}_{0} \in \mathbb{R}^{d}$, and $t_{0} \in \mathbb{R}$, there exists a unique $(\mathbf{u}, \mathbf{y}, \mathbf{x}) \in \mathcal{B}_{s}$ with $\mathbf{x}(t_{0}) = \mathbf{x}_{0}$, which is given by the variation of the constants formula: $\mathbf{x}(t) = e^{A(t-t_{0})}\mathbf{x}_{0} + \smallint_{t_{0}}^{t}e^{A(t-\tau)}B\mathbf{u}(\tau) d\tau$ for all $t \geq t_{0}$; $\mathbf{x}(t) =e^{A(t-t_{0})}\mathbf{x}_{0} - \smallint_{t}^{t_{0}}e^{A(t-\tau)}B\mathbf{u}(\tau) d\tau$ for all $t < t_{0}$; and $\mathbf{y} = C\mathbf{x} + D\mathbf{u}$. 

\labitem{\ref{sec:sot}\arabic{muni}}{nl:pis5} Let $\mathcal{B}_{s}$ be as in (\ref{eq:bne1}). We call the pair $(C,A)$ \emph{observable} if $(\mathbf{u}, \mathbf{y}, \mathbf{x}), (\mathbf{u}, \mathbf{y}, \hat{\mathbf{x}}) \in \mathcal{B}_{s}$ imply $\mathbf{x} = \hat{\mathbf{x}}$ \citep[Definition 5.3.2]{JWIMTSC}. With the notation $V_{o} \coloneqq \text{col}(C \hspace{0.15cm} CA \hspace{0.15cm} \cdots \hspace{0.15cm}  CA^{d-1})$, then $(C,A)$ is observable if and only if $\text{rank}(V_{o}) = d$ \citep[Theorem 5.3.9]{JWIMTSC}. Now, let $\text{rank}(V_{o}) = d_{1} < d$; let the columns of $S_{2}$ be a basis for the set $\lbrace \mathbf{z} \in \mathbb{R}^{d} \mid V_{o}\mathbf{z} = 0\rbrace$; let $S = [S_{1} \hspace{0.15cm} S_{2}]$ be nonsingular; and partition $T \coloneqq S^{-1}$ compatibly with $S$ as $T = \text{col}(T_{1} \hspace{0.15cm} T_{2})$. Then
\begin{equation*}
\begin{bmatrix}T_{1} \\ T_{2}\end{bmatrix}A\begin{bmatrix}S_{1}& S_{2}\end{bmatrix} \eqqcolon \begin{bmatrix}\tilde{A}_{11}& 0 \\\tilde{A}_{21}& \tilde{A}_{22}\end{bmatrix},
\hspace{0.1cm} C\begin{bmatrix}S_{1}& S_{2}\end{bmatrix} \eqqcolon \begin{bmatrix}\tilde{C}_{1}& 0\end{bmatrix},
\end{equation*}
where $(\tilde{C}_{1}, \tilde{A}_{11})$ is observable: the \emph{observer staircase form} \citep[Corollary 5.3.14]{JWIMTSC}.

\labitem{\ref{sec:sot}\arabic{muni}}{nl:pis6} Let $\mathcal{B}_{s}$ be as in (\ref{eq:bne1}); let $V_{o}$, $S$ and $T$ be as in note \ref{nl:pis5}; let $\mathcal{B} \coloneqq \mathcal{B}_{s}^{(\mathbf{u}, \mathbf{y})}$; and let $t_{0} \in \mathbb{R}$. If $(\mathbf{u}, \mathbf{y}, \mathbf{x}) \in \mathcal{B}_{s}$, $(\hat{\mathbf{u}}, \hat{\mathbf{y}}) \in \mathcal{B}$, and $(\hat{\mathbf{u}}(t),\hat{\mathbf{y}}(t)) = (\mathbf{u}(t), \mathbf{y}(t))$ for all $t < t_{0}$, then there exists $(\hat{\mathbf{u}}, \hat{\mathbf{y}}, \hat{\mathbf{x}}) \in \mathcal{B}_{s}$ with $\hat{\mathbf{x}}(t_{0}) = \mathbf{x}(t_{0})$. This follows from the following two observations, which are easily shown from the variation of the constants formula: (i) if $(\mathbf{u}, \mathbf{y}) \in \mathcal{B}$ and $\mathbf{z} \in \mathbb{R}^{d-d_{1}}$, then there exists $(\mathbf{u}, \mathbf{y}, \mathbf{x}) \in \mathcal{B}_{s}$ with $T_{2}\mathbf{x}(t_{0}) = \mathbf{z}$; and (ii) if $(\mathbf{u}, \mathbf{y}, \mathbf{x}), (\hat{\mathbf{u}}, \hat{\mathbf{y}}, \hat{\mathbf{x}}) \in \mathcal{B}_{s}$, and $(\mathbf{u}(t), \mathbf{y}(t)) = (\hat{\mathbf{u}}(t), \hat{\mathbf{y}}(t))$ for all $t < t_{0}$, then $T_{1}\mathbf{x}(t_{0}) = T_{1}\hat{\mathbf{x}}(t_{0})$.

Also, with $\tilde{A}_{11}$ and $\tilde{C}_{1}$ as in note \ref{nl:pis5}, and $\tilde{B}_{1} \coloneqq T_{1}B$, then it follows from the variation of the constants formula that $\mathcal{B} =  \tilde{\mathcal{B}}_{s}^{(\mathbf{u}, \mathbf{y})}$, with
\begin{align*}
&\hspace*{-0.1cm} \tilde{\mathcal{B}}_{s} {=} \lbrace (\mathbf{u}, \mathbf{y}, \tilde{\mathbf{x}}) \in \mathcal{L}_{1}^{\text{loc}}\left(\mathbb{R}, \mathbb{R}^{n}\right) {\times} \mathcal{L}_{1}^{\text{loc}}\left(\mathbb{R}, \mathbb{R}^{n}\right) {\times} \mathcal{L}_{1}^{\text{loc}}\left(\mathbb{R}, \mathbb{R}^{d_{1}}\right) \\
& \hspace{1.0cm} \text{such that } \tfrac{d\tilde{\mathbf{x}}}{dt} = \tilde{A}_{11}\tilde{\mathbf{x}} {+} \tilde{B}_{1}\mathbf{u} \text{ and } \mathbf{y} = \tilde{C}_{1}\tilde{\mathbf{x}} {+} D\mathbf{u}\rbrace.
\end{align*}

\labitem{\ref{sec:sot}\arabic{muni}}{nl:pis8} Let $\hat{\mathcal{B}} = \lbrace (\mathbf{u}, \mathbf{x}) \in \mathcal{L}_{1}^{\text{loc}}\left(\mathbb{R}, \mathbb{R}^{n}\right) \times \mathcal{L}_{1}^{\text{loc}}\left(\mathbb{R}, \mathbb{R}^{d}\right) \mid \tfrac{d\mathbf{x}}{dt} = A\mathbf{x} + B\mathbf{u}\rbrace$ and let $V_{c} \coloneqq [B \hspace{0.12cm} AB \hspace{0.12cm} \cdots \hspace{0.12cm} A^{d-1}B]$. If $\mathcal{B}$ is controllable, then we also call the pair $(A,B)$ controllable. From \citep[Section 5.2.1]{JWIMTSC}, the following are equivalent: (i) $(A, B)$ is controllable; (ii) $[\lambda I {-} A \hspace{0.15cm} B]$ has full row rank for all $\lambda \in \mathbb{C}$; and (iii) $\text{rank}(V_{c}) = d$. Now, let $\mathcal{B}_{s}$ be as in (\ref{eq:bne1}); let $\mathcal{B} {\coloneqq} \mathcal{B}_{s}^{(\mathbf{u}, \mathbf{y})}$; and let $(C,A)$ be observable. Then $\mathcal{B}$ is controllable (resp., stabilizable) if and only if $[\lambda I {-} A \hspace{0.15cm} B]$ has full row rank for all $\lambda \in \mathbb{C}$ (resp., $\lambda \in \overbar{\mathbb{C}}_{+}$). The proof is similar to \citep[proof of Theorem 5.2]{THBRSF}.

\end{remunerate}

\bibliographystyle{ifacconf-harvard}        
\bibliography{passive_behav_refs}           



\begin{wrapfigure}{L}{0.13\textwidth}
\centering
\includegraphics[width=0.15\textwidth]{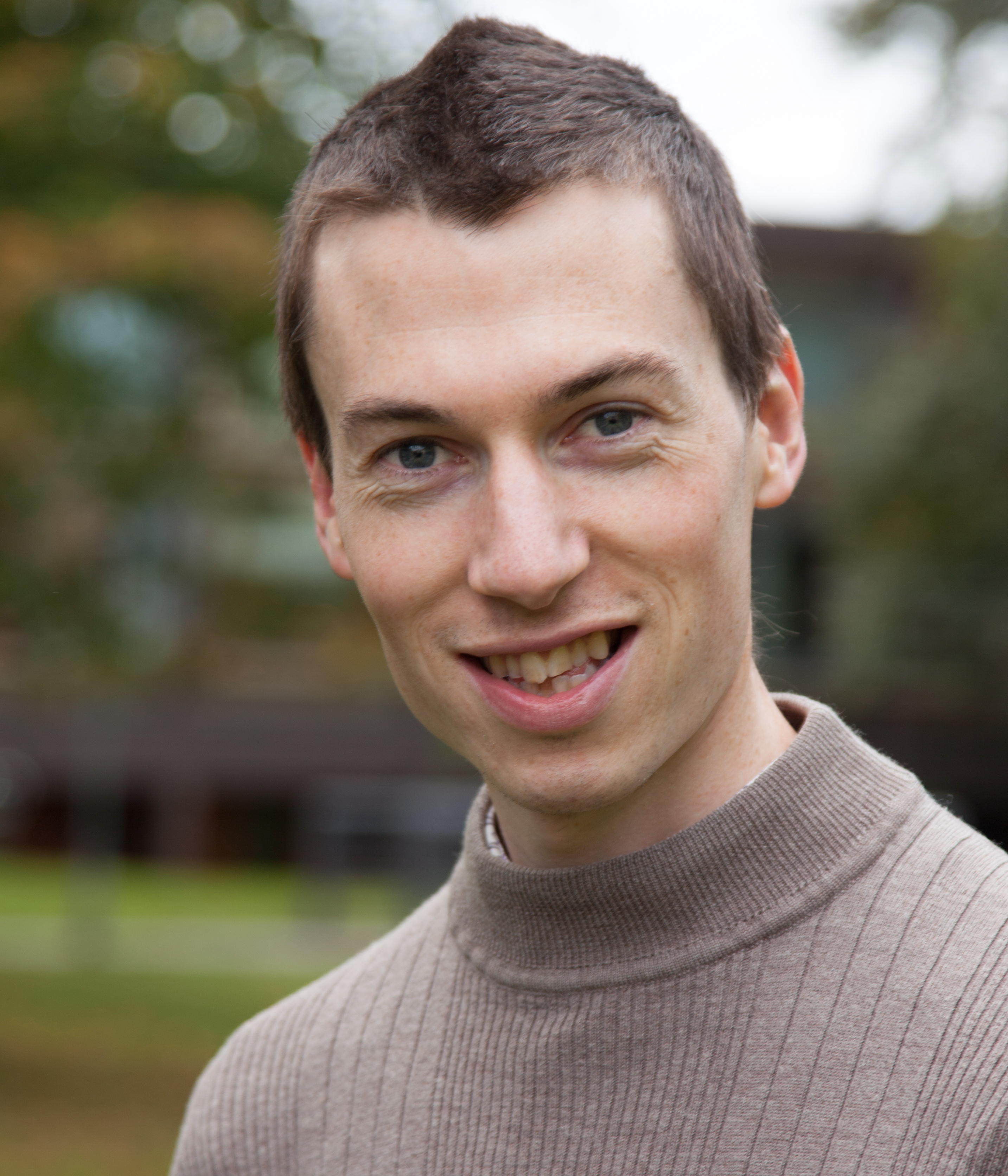}
\end{wrapfigure}

$\textbf{Timothy H.\ Hughes}$ received the M.Eng. degree in mechanical engineering, and the Ph.D. degree in control engineering, from the University of Cambridge, U.K., in 2007 and 2014, respectively.

From 2007 to 2010 he was employed as a Mechanical Engineer at The Technology Partnership, Hertfordshire, U.K; and from 2013 to 2017 he held a Research Fellowship at the University of Cambridge. He is now a Lecturer at the Department of Mathematics at the University of Exeter.

\vfill

\end{document}